\documentclass[
  reprint,
  amsmath,amssymb,
  pra,
  aps
]{revtex4-2}

\usepackage{graphicx}
\usepackage{bm}
\usepackage{booktabs}
\usepackage{xcolor}
\usepackage{hyperref}
\hypersetup{colorlinks=true,linkcolor=blue,citecolor=blue,urlcolor=blue}

\newcommand{\F}{\mathbb{F}}
\newcommand{\bfH}{\mathbf{H}}
\newcommand{\bfA}{\mathbf{A}}
\newcommand{\bfp}{\bm{p}}
\newcommand{\bfe}{\bm{e}}
\newcommand{\bfs}{\bm{s}}
\newcommand{\ehat}{\hat{\bfe}}
\DeclareMathOperator{\sign}{sign}

\newcommand{\PriorTilt}{\textsc{PriorTilt}}

\begin{document}

\title{PriorTilt: Decoding quantum LDPC codes with sparse perturbations of belief propagation priors}

\author{Min Ye, Nicolas Delfosse}
\affiliation{IonQ Inc.}

\date{\today}

\begin{abstract}
Although belief propagation (BP) is a widely adopted decoder for classical low-density parity-check (LDPC) codes, its accuracy is limited in the quantum setting. We introduce \PriorTilt{}, a variant of BP obtained by selectively perturbing a small fraction of its prior log-likelihood ratios using offsets determined by recent posterior and outgoing-message statistics, while preserving the same message-passing rules. We compare \PriorTilt{} with Relay-BP and beam search decoders for circuit-level decoding of three benchmark codes. \PriorTilt{} achieves a lower logical error rate (LER) than both baselines at every tested data point. For the $[[144,12,12]]$ Gross code BB144 at noise rate $p=10^{-3}$, the LER reductions reach $4.3\times$ relative to Relay-BP and $3.1\times$ relative to the beam search decoder. At $p=10^{-3}$, \PriorTilt{} has lower mean decoding times than Relay-BP on all three codes and achieves mean decoding speedups of approximately $2.6\times$--$3.0\times$ over the beam search decoder. For BB144 at $p=10^{-3}$, \PriorTilt{} achieves an LER approximately $610\times$ lower than the standard min-sum BP decoder with the same iteration budget, while perturbing fewer than $0.5\%$ of its priors per round on average.
\end{abstract}

\maketitle

\section{Introduction}
\label{sec:introduction}

The success of classical low-density parity-check (LDPC) codes stems from the combination of sparse parity checks and efficient iterative decoding~\cite{gallager1962low,mackay2003information,richardson2008modern}. Belief propagation (BP) is central to this approach~\cite{pearl1982reverend}. Quantum LDPC codes~\cite{mackay2004sparse} offer a route to reducing the physical-qubit overhead of fault-tolerant quantum computing. Their development includes hypergraph-product and related two-block constructions~\cite{tillich2013quantum,kovalev2013quantum}, balanced-product codes~\cite{breuckmann2021balanced}, and asymptotically good code families~\cite{panteleev2022asymptotically,leverrier2022quantum}. Circuit-level studies have demonstrated promising overhead reductions for several code families and hardware architectures~\cite{higgott2024constructions,bravyi2024high,ye2025quantum,scruby2026high,tremblay2022constant,aydin2025cyclic}. Realizing these gains requires decoders that combine accurate error correction with sufficiently low latency~\cite{terhal2015quantum,battistel2023real,khalid2025impacts}.

BP uses computationally efficient local updates on the Tanner graph~\cite{tanner1981recursive}, but its effectiveness depends on the graph structure, particularly its short cycles~\cite{hu2001progressive}. Quantum decoding introduces further difficulties: stabilizer degeneracy groups distinct errors into equivalent corrections~\cite{gottesman1997stabilizer}, while short cycles and symmetries can cause BP to oscillate or fail to find a syndrome-consistent solution~\cite{poulin2008iterative}. Combining BP with ordered statistics decoding (BP-OSD) substantially improves accuracy~\cite{panteleev2021degenerate,roffe_decoding_2020,Roffe_LDPC_Python_tools_2022}, but the associated Gaussian elimination can be costly. Linear-algebra costs also arise in generalized union-find decoding~\cite{delfosse2022toward}. Ambiguity clustering~\cite{wolanski2024introducing} and localized statistics decoding~\cite{hillmann2024localized} reduce this burden by organizing corrections into smaller subproblems.

Other approaches improve message passing through stabilizer inactivation~\cite{du2022stabilizer}, guided decimation~\cite{yao2024belief,gong2024toward}, cycle-free post-processing~\cite{iolius2024almost}, automorphism ensembles~\cite{koutsioumpas2025automorphism}, parallel trials~\cite{wang2026fully}, and updates designed to resolve trapping sets~\cite{pradhan2026fast}. Search and optimization methods provide complementary strategies, including decision trees~\cite{ott2025decision}, primal-dual optimization~\cite{wu2025minimum}, integer programming and $A^*$ search~\cite{beni2025tesseract}, and branch-and-bound-inspired restarts~\cite{valentini2026restart}. The Frontier decoder~\cite{leverrier2026approximating} uses pruned dynamic programming: it merges partial error configurations with the same residual syndrome and logical label, then prunes the resulting set to limit its size.

Two recent developments are particularly relevant to this work. Relay-BP~\cite{muller2025improved} chains BP runs with spatially disordered memory strengths, improving convergence and decoding accuracy while retaining local message passing. Its practical potential has motivated FPGA implementations and hardware-aware comparisons~\cite{maurer2025real,maurya2025fpga}, as well as GPU implementations combined with OSD post-processing~\cite{markidis2026engineering}. Relay-BP has also been used to evaluate logical operations in a modular bivariate bicycle architecture~\cite{yoder2025tour} and new balanced-product codes~\cite{hong2026quantum}. Independent work on graph transformations for correlated-error decoding uses XYZ Relay-BP as an accuracy benchmark~\cite{maan2026decoding}. These applications and comparisons establish Relay-BP as an important reference for practical quantum LDPC decoding.

The beam search decoder~\cite{ye2026beam} combines BP with a bounded search over candidate error configurations. It has been applied to trapped-ion memory experiments~\cite{tham2026breakeven} and simulations beyond Pauli stochastic noise~\cite{hines2026simulating}, while streaming implementations support the walking cat architecture~\cite{tripier2026fault} and real-time decoding of large-scale applications on a single CPU~\cite{ye2026real}. Related developments include beam search with guided restarts~\cite{taghipour2026multistage} and a soft beam search decoder that extracts confidence scores from internal decoding data for post-selection, with experimental validation~\cite{aydin2026soft}.

Here we introduce \PriorTilt{}, a BP variant that modifies only the prior log-likelihood ratios while leaving the rest of the message-passing framework unchanged. After an initial ordinary-BP stage, the decoder applies randomized prior offsets in rounds of six iterations. Recent posterior and outgoing-message statistics identify error-prone nodes and determine the offset magnitudes. Each offset remains fixed within a round, and BP messages are retained throughout decoding. A three-phase parameter schedule adjusts the perturbation strength and directional bias across early, intermediate, and late rounds. The intuition is that, as easier instances converge, the remaining instances may require different perturbations to escape persistent BP failures. In our measurements, fewer than $0.5\%$ of error nodes receive nonzero offsets per modified-BP round on average. For circuit-level decoding of the $[[144,12,12]]$ gross code at physical error rate $p=0.001$, these sparse prior perturbations reduce the logical error rate by approximately $610\times$ relative to standard min-sum BP with the same maximum iteration budget.

We benchmark \PriorTilt{} against Relay-BP and beam search decoders under circuit-level noise on three codes: the bivariate bicycle codes BB144 $[[144,12,12]]$ and BB70 $[[70,6,9]]$, and the generalized bicycle code GB102 $[[102,22,9]]$~\cite{tripier2026fault}. For standard XZ decoding, \PriorTilt{} achieves lower logical error rates at every tested data point, with reductions reaching $4.3\times$ relative to Relay-BP and $3.1\times$ relative to beam search decoder on BB144. At $p=0.001$, \PriorTilt{} has lower mean decoding times than Relay-BP on all three codes and achieves mean decoding speedups of approximately $2.6\times$--$3.0\times$ over beam search. Its 99th-percentile decoding times are lower than those of beam search on all three codes and Relay-BP on two of the three codes. For XYZ decoding of BB144, \PriorTilt{} achieves essentially the same logical error rate as Relay-BP with a $1.77\times$ speedup. To facilitate reproducibility, the source code for \PriorTilt{} is available in~\cite{priortilt_github}.

\section{The decoding problem and the BP algorithm}
\label{sect:background}

Throughout this paper, we consider decoding under circuit-level noise. The decoding problem is specified by a parity-check matrix $\bfH\in\F_2^{M\times N}$, a logical operator matrix $\bfA\in\F_2^{K\times N}$, and error probabilities $\bfp=(p_0,\ldots,p_{N-1})$, where $M$ is the number of detectors, $N$ is the number of possible error mechanisms, and $K$ is the number of logical observables. The columns of both matrices label the error mechanisms. Entry $\bfH_{ij}=1$ if mechanism $j$ flips detector $i$, while $\bfA_{kj}=1$ if it flips logical observable $k$. For an unknown error vector $\bfe\in\F_2^N$, the observed syndrome is $\bfs=\bfH\bfe$. Given $\bfs$, the decoder returns an estimate $\ehat\in\F_2^N$. Successful decoding requires
$$
 \bfH\ehat=\bfs,\qquad \bfA\ehat=\bfA\bfe,
$$
with binary arithmetic over $\F_2$. The first condition can be checked by the decoder, and the second one determines logical success.

Min-sum belief propagation (BP) operates on the Tanner graph of $\bfH$. An error node $e_j$ and a detector node $d_i$ are connected whenever $\bfH_{ij}=1$. We use $\mathcal N(v)$ to denote the neighbors of node $v$. The input syndrome is $\bfs=(s_0,\ldots,s_{M-1})$, where $s_i$ is the component associated with detector $d_i$. The detector-to-error and error-to-detector messages at iteration $t$ are written as $D_{i\to j}(t)$ and $E_{j\to i}(t)$. Each error node has prior log-likelihood ratio (LLR)
$$
 \Lambda_j=\log\frac{1-p_j}{p_j},\qquad
 E_{j\to i}(0)=\Lambda_j.
$$
For $t\geq1$, an iteration first updates all detector messages,
\begin{align*}
 D_{i\to j}(t)={}&(-1)^{s_i}
 \prod_{e_{j'}\in\mathcal N(d_i)\setminus\{e_j\}}
 \sign\!\left(E_{j'\to i}(t-1)\right) \\
 &{}\cdot\min_{e_{j'}\in\mathcal N(d_i)\setminus\{e_j\}}
 \left|E_{j'\to i}(t-1)\right|,
\end{align*}
then updates the error messages and posterior LLRs,
\begin{align}
 E_{j\to i}(t)&=\Lambda_j+
 \sum_{d_{i'}\in\mathcal N(e_j)\setminus\{d_i\}}D_{i'\to j}(t),
 \label{eq:error_to_detector}\\
 \Lambda_j(t)&=\Lambda_j+
 \sum_{d_{i'}\in\mathcal N(e_j)}D_{i'\to j}(t).
 \label{eq:posterior_LLR}
\end{align}
We use the unmodified min-sum rule without normalization or damping, with $\sign(x)=+1$ for $x>0$ and $-1$ otherwise. Positive LLRs favor no error, and the hard decision is 
$$
\hat e_j(t)=\mathbf 1[\Lambda_j(t)\leq0].
$$ 
BP stops when $\bfH\ehat(t)=\bfs$ or its iteration budget is exhausted.

\section{PriorTilt decoder}
\label{sec:decoder}

\PriorTilt{} modifies only the prior LLRs in~\eqref{eq:error_to_detector} and \eqref{eq:posterior_LLR}, leaving the rest of BP unchanged. The decoder begins with up to $I_0=60$ ordinary BP iterations. If convergence has not been reached, it continues with modified BP in rounds of $r=6$ iterations each. The parameter $R$ specifies the maximum number of these rounds, giving a total budget of $I_0+rR$ BP iterations, including both ordinary and modified BP.

We use $t$ for the BP iteration index and $\tau(t)$ for the round index associated with iteration $t$:
$$
 \tau(t)=
 \begin{cases}
  0, & 1\leq t\leq I_0,\\
  \left\lceil\dfrac{t-I_0}{r}\right\rceil,
     & I_0<t\leq I_0+rR.
 \end{cases}
$$
Here $\tau(t)=0$ denotes the initial ordinary-BP interval. For each subsequent round $\tau\geq1$, the $r=6$ iterations $t=I_0+r(\tau-1)+1,\ldots,I_0+r\tau$ share the same round index.

Recall that $\Lambda_j$ without an iteration argument denotes the prior LLR, whereas $\Lambda_j(t)$ denotes the posterior LLR at iteration $t$. The \PriorTilt{} decoder replaces $\Lambda_j$ on the right-hand sides of Eqs.~\eqref{eq:error_to_detector} and \eqref{eq:posterior_LLR} with the effective prior
\begin{equation}
 \widetilde{\Lambda}_j(t)
 =\Lambda_j+o_j(\tau(t)),
 \label{eq:effective}
\end{equation}
where $o_j(\tau(t))$ is the prior offset assigned to all six iterations in round $\tau(t)$. We set $o_j(0)=0$ to include the initial ordinary-BP interval in the same formulation. For each error node $j$, all six iterations within a round share the same offset $o_j(\tau(t))$ and hence the same effective prior $\widetilde{\Lambda}_j(t)$, since they have the same round index $\tau(t)$.

In both the ordinary and modified BP stages, decoding stops as soon as the hard decision satisfies the syndrome equation. If the iteration budget is exhausted, the final hard decision is returned with a non-convergence flag. Unlike the Relay-BP decoder \cite{muller2025improved}, messages in the \PriorTilt{} decoder are never reset between rounds.

It remains to specify how the prior offsets are computed. The offset $o_j(\tau)$ for error node $e_j$ in round $\tau$ is determined by the node's posterior LLRs and outgoing messages over the $r=6$ iterations in round $\tau-1$.

For each error node $e_j$, define
\begin{equation}
 S_j(\tau)
 =\sum_{t=I_0+r(\tau-1)+1}^{I_0+r\tau}\Lambda_j(t),
 \label{eq:sum}
\end{equation}
where $\Lambda_j(t)$ is the posterior LLR at iteration $t$. For $\tau\geq 1$, this sums the posterior LLRs over the six iterations of round $\tau$. For $\tau=0$, the sum covers the final six iterations of the initial ordinary-BP interval. Over the same iterations, define the set of posterior LLRs and outgoing error-to-detector messages
\begin{align*}
 \mathcal V_j(\tau)
 = \hspace{-4pt} \bigcup_{t=I_0+r(\tau-1)+1}^{I_0+r\tau}
   \bigl(\{\Lambda_j(t)\}
    \cup
   \{E_{j\to i}(t):d_i\in\mathcal N(e_j)\}\bigr)
\end{align*}
and the clipped extrema
\begin{align}
 M_j^{-}(\tau)&=\min\bigl\{0,\min\mathcal V_j(\tau)\bigr\},\nonumber\\
 M_j^{+}(\tau)&=\max\bigl\{0,\max\mathcal V_j(\tau)\bigr\}.
 \label{eq:extrema}
\end{align}
Here $M_j^{-}(\tau)$ is the most negative value in $\mathcal V_j(\tau)$, or zero if no negative value exists; $M_j^{+}(\tau)$ is the largest positive value, or zero if no positive value exists. Thus, $S_j(\tau)$ sums the posterior LLRs alone, whereas $M_j^{-}(\tau)$ and $M_j^{+}(\tau)$ capture the negative and positive extremes of both the posterior LLRs and outgoing messages. These statistics are updated on the fly during BP without storing the set $\mathcal V_j(\tau)$.

We now describe how these statistics determine the prior offset. For each error node $e_j$, we introduce a threshold $\theta\Lambda_j$, where $\theta>0$ is a tunable decoder parameter and $\Lambda_j$ is the prior LLR, which is typically positive. We call an error node $e_j$ \emph{error-prone} in round $\tau-1$ if
\begin{equation}
 S_j(\tau-1)<\theta\Lambda_j.
 \label{eq:trigger}
\end{equation}
This criterion includes nodes with strongly negative posterior LLR sums, for which BP favors the error hypothesis, as well as nodes with sums close to zero, for which BP does not clearly favor either the error or the no-error hypothesis. Only these error-prone nodes are eligible for a nonzero offset in round $\tau$. For all other nodes, i.e., nodes satisfying $S_j(\tau-1)\ge \theta\Lambda_j$, we set $o_j(\tau)=0$.

For each error-prone node, the offset is determined by three additional parameters: a probability $\pi$ and positive coefficients $c_1$ and $c_2$. Independently for each node and each round, we draw
\begin{equation}
 o_j(\tau)=
 \begin{cases}
  -c_1 M_j^{-}(\tau-1), & \text{with probability }\pi,\\
  -c_2 M_j^{+}(\tau-1), & \text{with probability }1-\pi.
 \end{cases}
 \label{eq:offset}
\end{equation}
The first choice produces a nonnegative offset, favoring the no-error hypothesis, while the second produces a nonpositive offset, favoring the error hypothesis. The probability $\pi$ is typically below $1/2$, so the latter choice is more frequent. The coefficients $c_1$ and $c_2$ scale the offsets according to the node's negative and positive extremes in the preceding round.

According to Eq.~\eqref{eq:effective}, the resulting offset $o_j(\tau)$ is applied throughout all six iterations in round $\tau$. Offsets are recomputed at every round boundary and replace, rather than accumulate with, the previous offsets. Each prior offset is therefore a finite-duration perturbation whose amplitude adapts to the ongoing BP trajectory.

To summarize the above description, the calculation of prior offsets depends on four tunable parameters,
$$
 \Theta=(\theta,\pi,c_1,c_2),
$$
where $\theta$ determines which nodes are error-prone through Eq.~\eqref{eq:trigger}, while $\pi$, $c_1$, and $c_2$ determine the offset through Eq.~\eqref{eq:offset}. In the \emph{uniform mode}, the parameter set $\Theta$ remains fixed across all rounds.

The \emph{three-phase mode} extends this construction by allowing the parameters to vary across three consecutive groups of rounds. The constant parameter set is replaced by a piecewise-constant schedule $\Theta(\tau)$. Specifically, let $0=b_0<b_1<b_2<b_3=R$ be integer round boundaries and define
$$
 \Theta(\tau)=\Theta_h=(\theta_h,\pi_h,c_{1,h},c_{2,h}),
$$
for $b_{h-1}<\tau\leq b_h$ and $h=1,2,3$. Changing phase selects a new parameter set without resetting BP messages or accumulated statistics. The initial ordinary-BP interval precedes these three phases. This mode has 14 tunable quantities: four parameters per phase and two internal boundaries, $b_1$ and $b_2$. The uniform mode is recovered when $\Theta_1=\Theta_2=\Theta_3$.

More generally, the same construction defines an $m$-phase mode for any $1\leq m\leq R$. The uniform mode corresponds to $m=1$, while $m=R$ allows an independently chosen parameter set for every round. In our tests, increasing the number of phases from one to two reduced the logical error rate, and a third phase provided a further reduction. A fourth phase yielded no clear additional improvement for any of the codes or circuits studied. We therefore use the three-phase mode for all simulations reported in this paper.

The 14 parameters of the three-phase mode are optimized for each circuit and physical error rate before estimating the corresponding logical error rate. Training pools contain only shots that remain unconverged after the $I_0$ ordinary BP iterations. Since the outcomes of shots that converge earlier are unaffected by the tunable parameters, this selection focuses optimization on the parameter-dependent part of the logical error rate. We construct separate pools for the $X$- and $Z$-memory circuits and minimize their combined failure count using coordinate descent over all 14 parameters. Parameter changes are accepted when they reduce this count, and successive sweeps use progressively smaller steps until no further improvement is found.

\section{Comparison with existing decoders}
\label{sec:comparison}

\textbf{Relay-BP.}
\PriorTilt{} and Relay-BP~\cite{muller2025improved} both replace the original prior $\Lambda_j$ in Eqs.~\eqref{eq:error_to_detector} and \eqref{eq:posterior_LLR} with a shifted prior while preserving the min-sum message-passing structure. Their major differences are as follows.

1. \emph{Round length and initialization.}
\PriorTilt{} begins with 60 ordinary BP iterations, followed by rounds of six modified-BP iterations. Relay-BP begins with 80 memory-modified BP iterations, followed by legs of 60 iterations each. Thus, \PriorTilt{} uses much shorter rounds: its six-iteration round budget is one tenth of the budget for subsequent Relay-BP legs. The initial stages also differ: \PriorTilt{} starts with standard BP, whereas Relay-BP uses memory-modified BP from the outset.

2. \emph{Continuity of messages.}
Relay-BP resets the error-to-detector messages to the original prior LLRs at the beginning of each leg, while \PriorTilt{} retains the error-to-detector messages throughout decoding.

3. \emph{Offset update frequency.}
Relay-BP recomputes its prior offsets at every iteration. \PriorTilt{} updates its offsets only at round boundaries; for each error node, the offset remains fixed throughout the six iterations of a round.

4.  \emph{Selective prior modification.}
Relay-BP applies its memory update to every error node, typically producing nonzero prior offsets at almost all nodes. \PriorTilt{} permits nonzero offsets only for nodes satisfying Eq.~\eqref{eq:trigger}. As shown in Table~\ref{tab:offset_fraction}, the mean fraction of such nodes is below $0.5\%$ for all three benchmark codes at the tested physical error rates. This also bounds the mean fraction of nodes with nonzero offsets from above. Thus, \PriorTilt{} directly modifies the priors of only a small fraction of nodes, although the resulting changes can propagate throughout the graph through BP messages.

5. \emph{Offset construction and randomness.}
Written in our notation, the Relay-BP offset is
\begin{equation}
 o_j^{\mathrm{Relay}}(t)
 =\gamma_j\bigl[\Lambda_j(t-1)-\Lambda_j\bigr],
 \label{eq:relay_offset}
\end{equation}
where $\gamma_j$ is a memory strength held fixed within a leg and resampled for subsequent legs. This offset depends on the preceding posterior LLR. The \PriorTilt{} offset instead uses extrema of both posterior LLRs and outgoing messages over the preceding six iterations. Its sign is selected by a fresh random draw for each node satisfying Eq.~\eqref{eq:trigger} in each round, and its magnitude is determined by Eq.~\eqref{eq:offset}.

\begin{figure*}
    \centering
    \includegraphics[width=\linewidth]{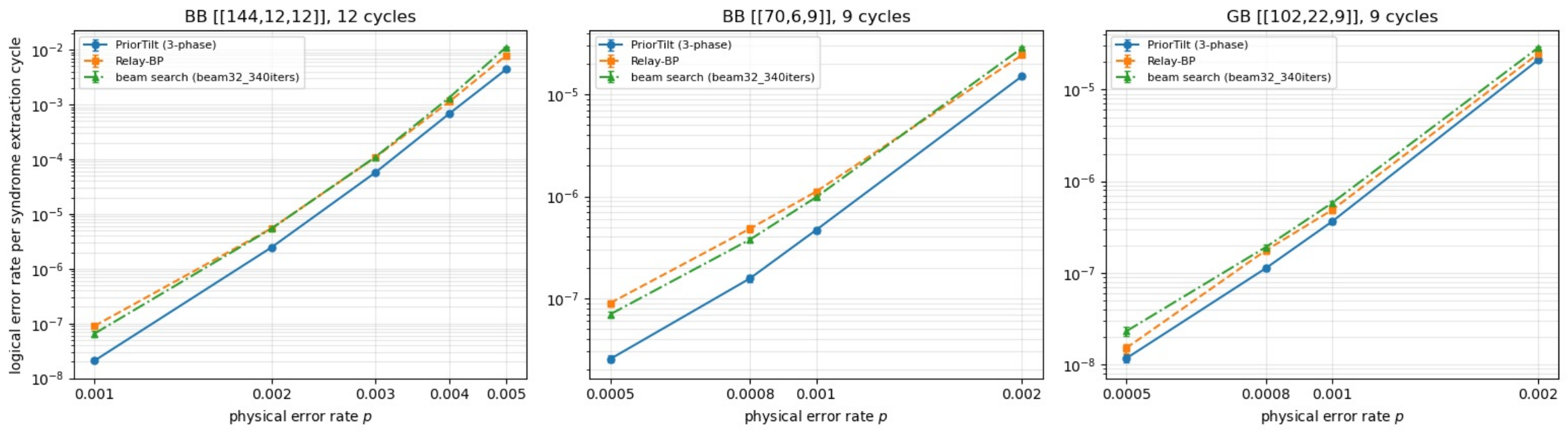}
    \caption{Logical error rates of \PriorTilt{}, Relay-BP, and beam search decoders for the three benchmark codes.}
    \label{fig:three_decoders_LER}
\end{figure*}

\textbf{BP-SF decoder.}
Both belief propagation with syndrome flipping (BP-SF)~\cite{wang2026fully} and \PriorTilt{} use BP statistics to identify nodes for targeted perturbations. BP-SF constructs a candidate set $\Phi$ from the nodes whose hard decisions change most frequently during an initial BP attempt. If that attempt fails, it samples trial error vectors supported on $\Phi$ and launches independent BP attempts on the corresponding perturbed syndromes. The candidate set $\Phi$ remains fixed throughout these additional attempts.

In contrast, \PriorTilt{} reevaluates every node's eligibility after each six-iteration round using the latest posterior statistics. Nodes excluded in earlier rounds can therefore become eligible later, allowing the perturbations to adapt to the evolving BP state. Moreover, \PriorTilt{} retains its BP messages throughout decoding and determines offset magnitudes from recent posterior and outgoing-message statistics, whereas BP-SF restarts BP and resets its messages at the start of each trial. The fixed set $\Phi$ restricts the support of BP-SF's trial perturbations, although its BP attempts can still correct errors outside that set.

For circuit-level decoding of BB144 at $p=0.001$, Fig.~7 of Ref.~\cite{wang2026fully} indicates that BP-SF reduces the LER by approximately $40$--$50\times$ relative to standard BP, with nearly equal maximum aggregate BP-iteration budgets of $10100$ and $10000$, respectively. Our simulations of BB144 at $p=0.001$ show that \PriorTilt{} reduces the LER by approximately $610\times$ relative to standard min-sum BP, with both decoders using a maximum of $18060$ iterations. Thus, \PriorTilt{} achieves a substantially larger LER reduction than BP-SF when each decoder is compared with standard BP at (nearly) the same maximum iteration budget.

An advantage of BP-SF is that its independent trials can run concurrently, as demonstrated by the GPU implementation in Ref.~\cite{wang2026gpu}. Successive \PriorTilt{} rounds, in contrast, depend on statistics from the preceding round, although message updates within each iteration remain amenable to parallel implementation.

\section{Simulation results}
\label{sec:simulation}

Our memory benchmarks comprise the bivariate bicycle codes BB144 $[[144,12,12]]$ and BB70 $[[70,6,9]]$, and the generalized bicycle code GB102 $[[102,22,9]]$. Each experiment uses $d$ syndrome extraction cycles, where $d$ is the code distance. For BB144, we use the circuits and uniform circuit-level noise model from Ref.~\cite{bravyi2024high}. For BB70 and GB102, we use the circuits from Ref.~\cite{tripier2026fault} together with its moving-qubit circuit-level noise model, although we ignore qubit loss and leakage for simplicity. We estimate logical error rates by decoding Monte Carlo samples generated with Stim~\cite{gidney2021stim}. Throughout this paper, the reported logical error rate (LER) is the sum of the $X$-memory and $Z$-memory LERs evaluated at the same physical error rate.

We compare \PriorTilt{} with two baseline decoders: Relay-BP~\cite{muller2025improved} and the beam search decoder~\cite{ye2026beam}. Unless otherwise stated, \PriorTilt{} uses 60 ordinary BP iterations followed by $R=3000$ rounds of six modified-BP iterations, giving a total budget of $18060$ BP iterations. We use the Relay-BP-1 configuration of Ref.~\cite{muller2025improved}, which returns immediately upon finding the first syndrome-consistent solution. Its maximum budget comprises 301 legs: an initial leg of 80 memory-modified BP iterations and 300 subsequent legs of 60 iterations each, giving a total budget of $18080$ iterations. These configurations give \PriorTilt{} and Relay-BP nearly equal maximum BP-iteration budgets. For the beam search decoder, we use the \texttt{beam32\_340iters} configuration from Ref.~\cite{ye2026beam}.

All \PriorTilt{} simulations use the three-phase mode. Its 14 parameters are optimized separately for each code and each physical error rate, with the $X$- and $Z$-memory experiments sharing the same set of parameters at each code and error rate. The optimized parameters are listed in Table~\ref{tab:optimal_parameters}.

These optimized parameters show a consistent pattern: the final phase has the largest $c_2$ and smaller $\pi$ and $c_1$ than the initial phase. Thus, later rounds select the nonpositive-offset branch more frequently and use a larger coefficient to scale its magnitude, while reducing the frequency and scale of nonnegative offsets. This pattern suggests relatively gentle perturbations early in decoding and a stronger bias toward the error hypothesis at error-prone nodes in shots that remain unconverged after many rounds. The intermediate-phase settings and the threshold parameter $\theta$ exhibit more code-dependent behavior.

Relay-BP has three code-dependent parameters: $\gamma_{\mathrm{init}}$, $\gamma_{\mathrm{lb}}$, and $\gamma_{\mathrm{ub}}$. In the initial leg, the memory strength $\gamma_j$ in Eq.~\eqref{eq:relay_offset} is set to $\gamma_{\mathrm{init}}$ for every error node. At the start of each subsequent leg, the memory strengths are drawn uniformly from the interval $[\gamma_{\mathrm{lb}},\gamma_{\mathrm{ub}}]$, independently across error nodes and legs, and remain fixed within that leg. For BB144, we use the optimized values reported in Ref.~\cite{muller2025improved}; for BB70 and GB102, we optimize these parameters ourselves. The optimized values are listed in Table~\ref{tab:relay_parameters}. For each code, the same Relay-BP parameter set yielded the lowest logical error rate across all physical error rates examined in our tuning scans. We therefore use one Relay-BP parameter set per code, whereas the optimized \PriorTilt{} parameters depend on the physical error rate.

\begin{table}[t]
 \caption{Optimized Relay-BP parameters. For each code, these
 settings are used across all physical error rates studied.}
 \label{tab:relay_parameters}
 \centering
 \begin{ruledtabular}
 \begin{tabular}{lcc}
 Code & $\gamma_{\mathrm{init}}$
      & $[\gamma_{\mathrm{lb}},\gamma_{\mathrm{ub}}]$ \\
 \hline
 BB144 & $0.125$ & $[-0.24,\,0.66]$ \\
 BB70  & $0.125$ & $[-0.24,\,0.50]$ \\
 GB102 & $0.03$  & $[-0.30,\,0.45]$
 \end{tabular}
 \end{ruledtabular}
\end{table}

Figure~\ref{fig:three_decoders_LER} compares the logical error rates of the three decoders for the benchmark codes, using the configurations specified above. \PriorTilt{} achieves the lowest LER at every tested physical error rate for all three codes. For BB144, it reduces the LER by factors of $1.7\times$--$4.3\times$ relative to Relay-BP and $1.9\times$--$3.1\times$ relative to beam search. The corresponding reductions for BB70 are $1.6\times$--$3.5\times$ and $1.9\times$--$2.7\times$, respectively. The gains for GB102 are more modest but consistent, ranging from $1.2\times$ to $1.6\times$ relative to Relay-BP and from $1.4\times$ to $2.0\times$ relative to beam search.

We also compare the mean and 99th-percentile decoding times of the three decoders for $Z$-memory experiments at $p=0.001$. For each code, all three decoders process the same set of $100000$ sampled shots using the configurations specified above. We measure single-thread decoding time on a single core of a 2023 Apple M3 Pro laptop. The mean and 99th-percentile times are reported in Tables~\ref{tab:timing_p001} and \ref{tab:timing_p001_p99}, respectively. Each timing measurement covers one complete shot, comprising 12 syndrome extraction cycles for BB144 and nine cycles for BB70 and GB102.

\begin{table}
 \caption{Mean decoding time per shot for $Z$-memory experiments at $p=0.001$. All three decoders process the same $100000$ shots for each code.}
 \label{tab:timing_p001}
 \centering
 \begin{ruledtabular}
 \begin{tabular}{lccc}
 Code & \PriorTilt{} & Relay-BP & Beam search\\
 \hline
 BB144 & $0.907~\mathrm{ms}$ & $1.038~\mathrm{ms}$ & $2.377~\mathrm{ms}$\\
 BB70  & $0.281~\mathrm{ms}$ & $0.416~\mathrm{ms}$ & $0.829~\mathrm{ms}$\\
 GB102 & $0.803~\mathrm{ms}$ & $1.224~\mathrm{ms}$ & $2.219~\mathrm{ms}$\\
 \end{tabular}
 \end{ruledtabular}
\end{table}

\begin{table}
 \caption{99th-percentile decoding time per shot for $Z$-memory experiments at $p=0.001$, computed from the same $100000$ shots per code used in Table~\ref{tab:timing_p001}.}
 \label{tab:timing_p001_p99}
 \centering
 \begin{ruledtabular}
 \begin{tabular}{lccc}
 Code & \PriorTilt{} & Relay-BP & Beam search\\
 \hline
 BB144 & $4.52~\mathrm{ms}$ & $3.33~\mathrm{ms}$ & $7.86~\mathrm{ms}$\\
 BB70  & $1.23~\mathrm{ms}$ & $1.37~\mathrm{ms}$ & $2.79~\mathrm{ms}$\\
 GB102 & $2.53~\mathrm{ms}$ & $3.93~\mathrm{ms}$ & $5.85~\mathrm{ms}$\\
 \end{tabular}
 \end{ruledtabular}
\end{table}

\PriorTilt{} has a lower mean decoding time than Relay-BP on all three codes and achieves mean decoding speedups of approximately $2.6\times$--$3.0\times$ over the beam search decoder. Its 99th-percentile decoding times are also lower than those of beam search on all three codes, by factors of approximately $1.7\times$--$2.3\times$. Compared with Relay-BP, \PriorTilt{} has lower 99th-percentile decoding times on BB70 and GB102, but a higher value on BB144.

\textbf{XYZ decoding of BB144.}
All results presented above use standard XZ decoding, which decodes the $X$- and $Z$-error components separately using $Z$- and $X$-detector values, respectively. XYZ decoding instead uses both detector types jointly, allowing the decoder to exploit correlations between the two error components, particularly those arising from Pauli-$Y$ faults. However, the resulting Tanner graph typically contains many length-four cycles, which pose a challenge for BP-based decoders. Further discussion can be found in Refs.~\cite{muller2025improved,ye2026beam}. XYZ decoding therefore tests a decoder's ability to exploit additional error information despite these short cycles. For decoders that handle them effectively, XYZ decoding can substantially reduce the LER. Otherwise, the difficulties caused by short cycles can outweigh the benefit of the additional information, resulting in a higher LER than under standard XZ decoding.

We next compare \PriorTilt{} and Relay-BP on XYZ decoding of BB144 at physical error rate $p=0.003$. We omit the beam search decoder because, for this benchmark, its XYZ-decoding LER is much worse than both its own standard XZ-decoding LER and the XYZ-decoding LERs of the other two decoders.

For Relay-BP, we use the Relay-BP-5 configuration of~\cite{muller2025improved}, with a maximum of 601 legs: an initial leg of 80 iterations followed by 600 legs of 60 iterations each, giving a total budget of $36080$ iterations. Decoding terminates after collecting five syndrome-consistent solutions or exhausting this budget. The parameters $\gamma_{\mathrm{init}}$, $\gamma_{\mathrm{lb}}$, and $\gamma_{\mathrm{ub}}$ are the BB144 values listed in Table~\ref{tab:relay_parameters}. This is the same configuration used for BB144 XYZ decoding in~\cite{muller2025improved}.

For \PriorTilt{}, we use the three-phase mode and increase the maximum number of rounds from $R=3000$ to $R=6000$. With 60 initial ordinary BP iterations and six modified-BP iterations per round, the total budget is $36060$ iterations, nearly equal to that of Relay-BP-5. However, \PriorTilt{} still returns immediately upon finding the first syndrome-consistent solution. We set the phase boundaries to $(b_1,b_2)=(50,420)$ and use the parameter sets
$\Theta_1=(2.00,0.070,0.50,0.80)$,
$\Theta_2=(0.686,0.1372,0.50,0.784)$, and
$\Theta_3=(1.372,0.049,0.50,1.568)$, where
$\Theta_h=(\theta_h,\pi_h,c_{1,h},c_{2,h})$.
These parameter sets apply to rounds $1$--$50$, $51$--$420$, and $421$--$6000$, respectively.

The XYZ-decoding results are summarized in Table~\ref{tab:xyz_results}. Both decoders achieve substantially lower LERs than in standard XZ decoding, where the LER per cycle is $5.74\times 10^{-5}$ for \PriorTilt{} and $1.09\times 10^{-4}$ for Relay-BP-1. In the XYZ setting, \PriorTilt{} and Relay-BP-5 achieve comparable LERs, while \PriorTilt{} requires $43.5\%$ less decoding time on average, corresponding to a $1.77\times$ speedup.

\begin{table}
 \caption{XYZ-decoding performance for BB144 at $p=0.003$. The LER per cycle is the sum of the $X$- and $Z$-memory LERs. The mean decoding time is measured per $Z$-memory shot, comprising 12 syndrome extraction cycles.}
 \label{tab:xyz_results}
 \centering
 \small
 \renewcommand{\arraystretch}{1.2}
 \begin{ruledtabular}
 \begin{tabular}{lcc}
 Decoder & LER per cycle & Mean decoding time\\
 \hline
 \PriorTilt{} & $1.04\times10^{-5}$ & $601~\mathrm{ms}$\\
 Relay-BP-5   & $1.10\times10^{-5}$ & $1064~\mathrm{ms}$\\
 \end{tabular}
 \end{ruledtabular}
\end{table}

\begin{table}
 \caption{Mean percentage of error nodes satisfying \eqref{eq:trigger} in the three-phase mode under standard XZ decoding. These percentages bound the mean fraction of nonzero prior offsets from above.}
 \label{tab:offset_fraction}
 \centering
 \small
 \renewcommand{\arraystretch}{1.2}
 \begin{ruledtabular}
 \begin{tabular}{lcc}
 Code & $p=0.001$ & $p=0.002$\\
 \hline
 BB144 & $0.19\%$ & $0.33\%$\\
 BB70  & $0.16\%$ & $0.31\%$\\
 GB102 & $0.33\%$ & $0.26\%$\\
 \end{tabular}
 \end{ruledtabular}
\end{table}

\textbf{Fraction of nonzero prior offsets.}
We quantify the sparsity of the prior modifications in the three-phase mode under standard XZ decoding. Since only nodes satisfying the error-prone condition in Eq.~\eqref{eq:trigger} can receive nonzero offsets, the fraction of error-prone nodes provides an upper bound on the fraction with nonzero offsets. We sum the number of nodes satisfying \eqref{eq:trigger} over all modified-BP rounds and divide by $N$ times the total number of these rounds, where $N$ is the number of error nodes in the Tanner graph. The counts are pooled over the $X$- and $Z$-memory experiments. The initial ordinary-BP interval and shots that converge within it do not contribute to this statistic.

For each code and each memory basis, we sample $100000$ shots at $p=0.001$ and $30000$ shots at $p=0.002$. Table~\ref{tab:offset_fraction} reports the resulting mean percentages. In every tested setting, fewer than $0.5\%$ of error nodes are eligible for nonzero prior offsets on average. These sparse prior modifications can nevertheless substantially improve decoding performance. For BB144 at $p=0.001$, standard min-sum BP with a maximum budget of $18060$ iterations yields an LER of $1.27\times10^{-5}$, whereas \PriorTilt{} achieves $2.08\times10^{-8}$ with the same maximum iteration budget. Thus, \PriorTilt{} reduces the LER by approximately $610\times$ while directly modifying the priors of fewer than $0.5\%$ of error nodes per modified-BP round on average and leaving the rest of the message-passing framework unchanged.

\begin{table*}[t]
 \caption{Optimized parameters for the three-phase mode of \PriorTilt{} in standard XZ decoding. Round $\tau$ uses $\Theta_1$ for $1\leq \tau\leq b_1$, $\Theta_2$ for $b_1<\tau\leq b_2$, and $\Theta_3$ for $b_2<\tau\leq3000$. Every tuple is $(\theta,\pi,c_1,c_2)$.}
 \label{tab:optimal_parameters}
 \small
 \begin{ruledtabular}
 \begin{tabular}{lccccc}
 Code & $p$ & $(b_1,b_2)$ & $\Theta_1$ & $\Theta_2$ & $\Theta_3$\\
 \hline
 BB144 & 0.001 & $(100,800)$
 & $(1.40,0.274,0.50,0.48)$ & $(1.19,0.195,0.50,0.77)$ & $(2.57,0.098,0.46,2.45)$\\
 BB144 & 0.002 & $(100,800)$
 & $(1.40,0.196,0.50,0.56)$ & $(1.40,0.139,0.50,0.55)$ & $(2.18,0.098,0.33,2.45)$\\
 BB144 & 0.003 & $(120,1000)$
 & $(1.40,0.196,0.50,0.39)$ & $(1.40,0.118,0.50,0.56)$ & $(2.18,0.140,0.20,1.75)$\\
 BB144 & 0.004 & $(120,1000)$
 & $(1.40,0.137,0.59,0.39)$ & $(1.40,0.118,0.43,0.56)$ & $(1.85,0.115,0.20,1.75)$\\
 BB144 & 0.005 & $(70,1000)$
 & $(1.40,0.196,0.50,0.39)$ & $(1.65,0.083,0.41,0.48)$ & $(1.85,0.115,0.17,1.49)$\\
 \hline
 BB70 & 0.0005 & $(80,860)$
 & $(2.00,0.548,0.59,0.48)$ & $(2.00,0.549,0.25,1.03)$ & $(2.36,0.281,0.35,3.47)$\\
 BB70 & 0.0008 & $(80,860)$
 & $(2.00,0.644,0.59,0.48)$ & $(2.00,0.549,0.25,1.03)$ & $(2.36,0.201,0.25,2.48)$\\
 BB70 & 0.001 & $(80,860)$
 & $(2.00,0.546,0.59,0.56)$ & $(2.00,0.549,0.25,1.03)$ & $(2.36,0.201,0.25,2.48)$\\
 BB70 & 0.002 & $(100,1070)$
 & $(2.00,0.463,0.50,0.39)$ & $(2.00,0.392,0.25,0.74)$ & $(2.00,0.170,0.18,1.77)$\\
 \hline
 GB102 & 0.0005 & $(100,1560)$
 & $(1.01,0.700,0.59,0.39)$ & $(1.70,0.588,0.60,0.48)$ & $(1.40,0.301,0.28,1.04)$\\
 GB102 & 0.0008 & $(100,1560)$
 & $(1.01,0.700,0.59,0.39)$ & $(1.70,0.588,0.60,0.48)$ & $(1.40,0.301,0.28,1.04)$\\
 GB102 & 0.001 & $(100,1560)$
 & $(1.19,0.500,0.59,0.39)$ & $(1.70,0.588,0.43,0.48)$ & $(1.40,0.301,0.28,1.04)$\\
 GB102 & 0.002 & $(200,1560)$
 & $(0.98,0.421,0.59,0.27)$ & $(2.36,0.420,0.43,0.48)$ & $(1.40,0.255,0.20,1.49)$\\
 \end{tabular}
 \end{ruledtabular}
\end{table*}

\section{Conclusion}

In this work, we designed the \PriorTilt{} decoder for quantum LDPC codes and demonstrated through circuit-level simulation that it outperforms state-of-the-art decoders such as Relay-BP and beam search decoder.

All three decoders build on BP. Unlike Relay-BP, which perturbs almost all of the BP priors at each iteration and resets error-to-detector messages regularly between legs, \PriorTilt{} only perturbs a small fraction of the priors and does not reset messages throughout decoding. In our sparsity benchmarks, \PriorTilt{} perturbs fewer than $0.5\%$ of priors per modified-BP round on average, yet achieves a $610\times$ reduction in LER over standard BP. These results show that selective perturbations of a small fraction of priors can substantially improve decoding performance while preserving the standard BP message-passing structure. This proximity to standard BP may facilitate efficient hardware implementations of \PriorTilt{}.

\section*{Acknowledgment}

The authors thank John Gamble, Aharon Brodutch, Nolan Coble, Edwin Tham, Felix Tripier, Mark Webster for inspiring discussions and valuable feedback on this work.

\bibliography{references}

\begin{thebibliography}{58}%
\makeatletter
\providecommand \@ifxundefined [1]{%
 \@ifx{#1\undefined}
}%
\providecommand \@ifnum [1]{%
 \ifnum #1\expandafter \@firstoftwo
 \else \expandafter \@secondoftwo
 \fi
}%
\providecommand \@ifx [1]{%
 \ifx #1\expandafter \@firstoftwo
 \else \expandafter \@secondoftwo
 \fi
}%
\providecommand \natexlab [1]{#1}%
\providecommand \enquote  [1]{``#1''}%
\providecommand \bibnamefont  [1]{#1}%
\providecommand \bibfnamefont [1]{#1}%
\providecommand \citenamefont [1]{#1}%
\providecommand \href@noop [0]{\@secondoftwo}%
\providecommand \href [0]{\begingroup \@sanitize@url \@href}%
\providecommand \@href[1]{\@@startlink{#1}\@@href}%
\providecommand \@@href[1]{\endgroup#1\@@endlink}%
\providecommand \@sanitize@url [0]{\catcode `\\12\catcode `\$12\catcode `\&12\catcode `\#12\catcode `\^12\catcode `\_12\catcode `\%12\relax}%
\providecommand \@@startlink[1]{}%
\providecommand \@@endlink[0]{}%
\providecommand \url  [0]{\begingroup\@sanitize@url \@url }%
\providecommand \@url [1]{\endgroup\@href {#1}{\urlprefix }}%
\providecommand \urlprefix  [0]{URL }%
\providecommand \Eprint [0]{\href }%
\providecommand \doibase [0]{https://doi.org/}%
\providecommand \selectlanguage [0]{\@gobble}%
\providecommand \bibinfo  [0]{\@secondoftwo}%
\providecommand \bibfield  [0]{\@secondoftwo}%
\providecommand \translation [1]{[#1]}%
\providecommand \BibitemOpen [0]{}%
\providecommand \bibitemStop [0]{}%
\providecommand \bibitemNoStop [0]{.\EOS\space}%
\providecommand \EOS [0]{\spacefactor3000\relax}%
\providecommand \BibitemShut  [1]{\csname bibitem#1\endcsname}%
\let\auto@bib@innerbib\@empty
\bibitem [{\citenamefont {Gallager}(1962)}]{gallager1962low}%
  \BibitemOpen
  \bibfield  {author} {\bibinfo {author} {\bibfnamefont {R.}~\bibnamefont {Gallager}},\ }\bibfield  {title} {\bibinfo {title} {Low-density parity-check codes},\ }\href {https://doi.org/10.1109/TIT.1962.1057683} {\bibfield  {journal} {\bibinfo  {journal} {IRE Transactions on Information Theory}\ }\textbf {\bibinfo {volume} {8}},\ \bibinfo {pages} {21} (\bibinfo {year} {1962})}\BibitemShut {NoStop}%
\bibitem [{\citenamefont {MacKay}(2003)}]{mackay2003information}%
  \BibitemOpen
  \bibfield  {author} {\bibinfo {author} {\bibfnamefont {D.~J.}\ \bibnamefont {MacKay}},\ }\href@noop {} {\emph {\bibinfo {title} {Information theory, inference and learning algorithms}}}\ (\bibinfo  {publisher} {Cambridge university press},\ \bibinfo {year} {2003})\BibitemShut {NoStop}%
\bibitem [{\citenamefont {Richardson}\ and\ \citenamefont {Urbanke}(2008)}]{richardson2008modern}%
  \BibitemOpen
  \bibfield  {author} {\bibinfo {author} {\bibfnamefont {T.}~\bibnamefont {Richardson}}\ and\ \bibinfo {author} {\bibfnamefont {R.}~\bibnamefont {Urbanke}},\ }\href@noop {} {\emph {\bibinfo {title} {Modern coding theory}}}\ (\bibinfo  {publisher} {Cambridge university press},\ \bibinfo {year} {2008})\BibitemShut {NoStop}%
\bibitem [{\citenamefont {Pearl}(1982)}]{pearl1982reverend}%
  \BibitemOpen
  \bibfield  {author} {\bibinfo {author} {\bibfnamefont {J.}~\bibnamefont {Pearl}},\ }\bibfield  {title} {\bibinfo {title} {Reverend bayes on inference engines: a distributed hierarchical approach},\ }in\ \href@noop {} {\emph {\bibinfo {booktitle} {Proceedings of the Second AAAI Conference on Artificial Intelligence}}},\ \bibinfo {series and number} {AAAI'82}\ (\bibinfo  {publisher} {AAAI Press},\ \bibinfo {year} {1982})\ p.\ \bibinfo {pages} {133–136}\BibitemShut {NoStop}%
\bibitem [{\citenamefont {MacKay}\ \emph {et~al.}(2004)\citenamefont {MacKay}, \citenamefont {Mitchison},\ and\ \citenamefont {McFadden}}]{mackay2004sparse}%
  \BibitemOpen
  \bibfield  {author} {\bibinfo {author} {\bibfnamefont {D.~J.}\ \bibnamefont {MacKay}}, \bibinfo {author} {\bibfnamefont {G.}~\bibnamefont {Mitchison}},\ and\ \bibinfo {author} {\bibfnamefont {P.~L.}\ \bibnamefont {McFadden}},\ }\bibfield  {title} {\bibinfo {title} {Sparse-graph codes for quantum error correction},\ }\href@noop {} {\bibfield  {journal} {\bibinfo  {journal} {IEEE Transactions on Information Theory}\ }\textbf {\bibinfo {volume} {50}},\ \bibinfo {pages} {2315} (\bibinfo {year} {2004})}\BibitemShut {NoStop}%
\bibitem [{\citenamefont {Tillich}\ and\ \citenamefont {Z{\'e}mor}(2013)}]{tillich2013quantum}%
  \BibitemOpen
  \bibfield  {author} {\bibinfo {author} {\bibfnamefont {J.-P.}\ \bibnamefont {Tillich}}\ and\ \bibinfo {author} {\bibfnamefont {G.}~\bibnamefont {Z{\'e}mor}},\ }\bibfield  {title} {\bibinfo {title} {Quantum {LDPC} codes with positive rate and minimum distance proportional to the square root of the blocklength},\ }\href@noop {} {\bibfield  {journal} {\bibinfo  {journal} {IEEE Transactions on Information Theory}\ }\textbf {\bibinfo {volume} {60}},\ \bibinfo {pages} {1193} (\bibinfo {year} {2013})}\BibitemShut {NoStop}%
\bibitem [{\citenamefont {Kovalev}\ and\ \citenamefont {Pryadko}(2013)}]{kovalev2013quantum}%
  \BibitemOpen
  \bibfield  {author} {\bibinfo {author} {\bibfnamefont {A.~A.}\ \bibnamefont {Kovalev}}\ and\ \bibinfo {author} {\bibfnamefont {L.~P.}\ \bibnamefont {Pryadko}},\ }\bibfield  {title} {\bibinfo {title} {Quantum kronecker sum-product low-density parity-check codes with finite rate},\ }\href@noop {} {\bibfield  {journal} {\bibinfo  {journal} {Physical Review A—Atomic, Molecular, and Optical Physics}\ }\textbf {\bibinfo {volume} {88}},\ \bibinfo {pages} {012311} (\bibinfo {year} {2013})}\BibitemShut {NoStop}%
\bibitem [{\citenamefont {Breuckmann}\ and\ \citenamefont {Eberhardt}(2021)}]{breuckmann2021balanced}%
  \BibitemOpen
  \bibfield  {author} {\bibinfo {author} {\bibfnamefont {N.~P.}\ \bibnamefont {Breuckmann}}\ and\ \bibinfo {author} {\bibfnamefont {J.~N.}\ \bibnamefont {Eberhardt}},\ }\bibfield  {title} {\bibinfo {title} {Balanced product quantum codes},\ }\href@noop {} {\bibfield  {journal} {\bibinfo  {journal} {IEEE Transactions on Information Theory}\ }\textbf {\bibinfo {volume} {67}},\ \bibinfo {pages} {6653} (\bibinfo {year} {2021})}\BibitemShut {NoStop}%
\bibitem [{\citenamefont {Panteleev}\ and\ \citenamefont {Kalachev}(2022)}]{panteleev2022asymptotically}%
  \BibitemOpen
  \bibfield  {author} {\bibinfo {author} {\bibfnamefont {P.}~\bibnamefont {Panteleev}}\ and\ \bibinfo {author} {\bibfnamefont {G.}~\bibnamefont {Kalachev}},\ }\bibfield  {title} {\bibinfo {title} {Asymptotically good quantum and locally testable classical ldpc codes},\ }in\ \href@noop {} {\emph {\bibinfo {booktitle} {Proceedings of the 54th annual ACM SIGACT symposium on theory of computing}}}\ (\bibinfo {year} {2022})\ pp.\ \bibinfo {pages} {375--388}\BibitemShut {NoStop}%
\bibitem [{\citenamefont {Leverrier}\ and\ \citenamefont {Z{\'e}mor}(2022)}]{leverrier2022quantum}%
  \BibitemOpen
  \bibfield  {author} {\bibinfo {author} {\bibfnamefont {A.}~\bibnamefont {Leverrier}}\ and\ \bibinfo {author} {\bibfnamefont {G.}~\bibnamefont {Z{\'e}mor}},\ }\bibfield  {title} {\bibinfo {title} {Quantum tanner codes},\ }in\ \href@noop {} {\emph {\bibinfo {booktitle} {2022 IEEE 63rd Annual Symposium on Foundations of Computer Science (FOCS)}}}\ (\bibinfo {organization} {IEEE},\ \bibinfo {year} {2022})\ pp.\ \bibinfo {pages} {872--883}\BibitemShut {NoStop}%
\bibitem [{\citenamefont {Higgott}\ and\ \citenamefont {Breuckmann}(2024)}]{higgott2024constructions}%
  \BibitemOpen
  \bibfield  {author} {\bibinfo {author} {\bibfnamefont {O.}~\bibnamefont {Higgott}}\ and\ \bibinfo {author} {\bibfnamefont {N.~P.}\ \bibnamefont {Breuckmann}},\ }\bibfield  {title} {\bibinfo {title} {Constructions and performance of hyperbolic and semi-hyperbolic floquet codes},\ }\href@noop {} {\bibfield  {journal} {\bibinfo  {journal} {PRX Quantum}\ }\textbf {\bibinfo {volume} {5}},\ \bibinfo {pages} {040327} (\bibinfo {year} {2024})}\BibitemShut {NoStop}%
\bibitem [{\citenamefont {Bravyi}\ \emph {et~al.}(2024)\citenamefont {Bravyi}, \citenamefont {Cross}, \citenamefont {Gambetta}, \citenamefont {Maslov}, \citenamefont {Rall},\ and\ \citenamefont {Yoder}}]{bravyi2024high}%
  \BibitemOpen
  \bibfield  {author} {\bibinfo {author} {\bibfnamefont {S.}~\bibnamefont {Bravyi}}, \bibinfo {author} {\bibfnamefont {A.~W.}\ \bibnamefont {Cross}}, \bibinfo {author} {\bibfnamefont {J.~M.}\ \bibnamefont {Gambetta}}, \bibinfo {author} {\bibfnamefont {D.}~\bibnamefont {Maslov}}, \bibinfo {author} {\bibfnamefont {P.}~\bibnamefont {Rall}},\ and\ \bibinfo {author} {\bibfnamefont {T.~J.}\ \bibnamefont {Yoder}},\ }\bibfield  {title} {\bibinfo {title} {High-threshold and low-overhead fault-tolerant quantum memory},\ }\href@noop {} {\bibfield  {journal} {\bibinfo  {journal} {Nature}\ }\textbf {\bibinfo {volume} {627}},\ \bibinfo {pages} {778} (\bibinfo {year} {2024})}\BibitemShut {NoStop}%
\bibitem [{\citenamefont {Ye}\ and\ \citenamefont {Delfosse}(2025)}]{ye2025quantum}%
  \BibitemOpen
  \bibfield  {author} {\bibinfo {author} {\bibfnamefont {M.}~\bibnamefont {Ye}}\ and\ \bibinfo {author} {\bibfnamefont {N.}~\bibnamefont {Delfosse}},\ }\bibfield  {title} {\bibinfo {title} {Quantum error correction for long chains of trapped ions},\ }\href@noop {} {\bibfield  {journal} {\bibinfo  {journal} {Quantum}\ }\textbf {\bibinfo {volume} {9}},\ \bibinfo {pages} {1920} (\bibinfo {year} {2025})}\BibitemShut {NoStop}%
\bibitem [{\citenamefont {Scruby}\ \emph {et~al.}(2026)\citenamefont {Scruby}, \citenamefont {Hillmann},\ and\ \citenamefont {Roffe}}]{scruby2026high}%
  \BibitemOpen
  \bibfield  {author} {\bibinfo {author} {\bibfnamefont {T.~R.}\ \bibnamefont {Scruby}}, \bibinfo {author} {\bibfnamefont {T.}~\bibnamefont {Hillmann}},\ and\ \bibinfo {author} {\bibfnamefont {J.}~\bibnamefont {Roffe}},\ }\bibfield  {title} {\bibinfo {title} {High-threshold, low-overhead and single-shot decodable fault-tolerant quantum memory},\ }\href@noop {} {\bibfield  {journal} {\bibinfo  {journal} {PRX Quantum}\ }\textbf {\bibinfo {volume} {7}},\ \bibinfo {pages} {020310} (\bibinfo {year} {2026})}\BibitemShut {NoStop}%
\bibitem [{\citenamefont {Tremblay}\ \emph {et~al.}(2022)\citenamefont {Tremblay}, \citenamefont {Delfosse},\ and\ \citenamefont {Beverland}}]{tremblay2022constant}%
  \BibitemOpen
  \bibfield  {author} {\bibinfo {author} {\bibfnamefont {M.~A.}\ \bibnamefont {Tremblay}}, \bibinfo {author} {\bibfnamefont {N.}~\bibnamefont {Delfosse}},\ and\ \bibinfo {author} {\bibfnamefont {M.~E.}\ \bibnamefont {Beverland}},\ }\bibfield  {title} {\bibinfo {title} {Constant-overhead quantum error correction with thin planar connectivity},\ }\href@noop {} {\bibfield  {journal} {\bibinfo  {journal} {Physical Review Letters}\ }\textbf {\bibinfo {volume} {129}},\ \bibinfo {pages} {050504} (\bibinfo {year} {2022})}\BibitemShut {NoStop}%
\bibitem [{\citenamefont {Aydin}\ \emph {et~al.}(2025)\citenamefont {Aydin}, \citenamefont {Delfosse},\ and\ \citenamefont {Tham}}]{aydin2025cyclic}%
  \BibitemOpen
  \bibfield  {author} {\bibinfo {author} {\bibfnamefont {A.}~\bibnamefont {Aydin}}, \bibinfo {author} {\bibfnamefont {N.}~\bibnamefont {Delfosse}},\ and\ \bibinfo {author} {\bibfnamefont {E.}~\bibnamefont {Tham}},\ }\bibfield  {title} {\bibinfo {title} {Cyclic hypergraph product code},\ }\href@noop {} {\bibfield  {journal} {\bibinfo  {journal} {arXiv preprint arXiv:2511.09683}\ } (\bibinfo {year} {2025})}\BibitemShut {NoStop}%
\bibitem [{\citenamefont {Terhal}(2015)}]{terhal2015quantum}%
  \BibitemOpen
  \bibfield  {author} {\bibinfo {author} {\bibfnamefont {B.~M.}\ \bibnamefont {Terhal}},\ }\bibfield  {title} {\bibinfo {title} {Quantum error correction for quantum memories},\ }\href@noop {} {\bibfield  {journal} {\bibinfo  {journal} {Reviews of Modern Physics}\ }\textbf {\bibinfo {volume} {87}},\ \bibinfo {pages} {307} (\bibinfo {year} {2015})}\BibitemShut {NoStop}%
\bibitem [{\citenamefont {Battistel}\ \emph {et~al.}(2023)\citenamefont {Battistel}, \citenamefont {Chamberland}, \citenamefont {Johar}, \citenamefont {Overwater}, \citenamefont {Sebastiano}, \citenamefont {Skoric}, \citenamefont {Ueno},\ and\ \citenamefont {Usman}}]{battistel2023real}%
  \BibitemOpen
  \bibfield  {author} {\bibinfo {author} {\bibfnamefont {F.}~\bibnamefont {Battistel}}, \bibinfo {author} {\bibfnamefont {C.}~\bibnamefont {Chamberland}}, \bibinfo {author} {\bibfnamefont {K.}~\bibnamefont {Johar}}, \bibinfo {author} {\bibfnamefont {R.~W.}\ \bibnamefont {Overwater}}, \bibinfo {author} {\bibfnamefont {F.}~\bibnamefont {Sebastiano}}, \bibinfo {author} {\bibfnamefont {L.}~\bibnamefont {Skoric}}, \bibinfo {author} {\bibfnamefont {Y.}~\bibnamefont {Ueno}},\ and\ \bibinfo {author} {\bibfnamefont {M.}~\bibnamefont {Usman}},\ }\bibfield  {title} {\bibinfo {title} {Real-time decoding for fault-tolerant quantum computing: Progress, challenges and outlook},\ }\href@noop {} {\bibfield  {journal} {\bibinfo  {journal} {Nano Futures}\ }\textbf {\bibinfo {volume} {7}},\ \bibinfo {pages} {032003} (\bibinfo {year} {2023})}\BibitemShut {NoStop}%
\bibitem [{\citenamefont {Khalid}\ \emph {et~al.}(2025)\citenamefont {Khalid}, \citenamefont {Silva}, \citenamefont {Dagnew}, \citenamefont {Dvir}, \citenamefont {Wertheim}, \citenamefont {Gruda}, \citenamefont {Kong}, \citenamefont {Kramer}, \citenamefont {Webb}, \citenamefont {Scherer} \emph {et~al.}}]{khalid2025impacts}%
  \BibitemOpen
  \bibfield  {author} {\bibinfo {author} {\bibfnamefont {A.}~\bibnamefont {Khalid}}, \bibinfo {author} {\bibfnamefont {A.}~\bibnamefont {Silva}}, \bibinfo {author} {\bibfnamefont {G.~A.}\ \bibnamefont {Dagnew}}, \bibinfo {author} {\bibfnamefont {T.}~\bibnamefont {Dvir}}, \bibinfo {author} {\bibfnamefont {O.}~\bibnamefont {Wertheim}}, \bibinfo {author} {\bibfnamefont {M.}~\bibnamefont {Gruda}}, \bibinfo {author} {\bibfnamefont {X.}~\bibnamefont {Kong}}, \bibinfo {author} {\bibfnamefont {M.}~\bibnamefont {Kramer}}, \bibinfo {author} {\bibfnamefont {Z.}~\bibnamefont {Webb}}, \bibinfo {author} {\bibfnamefont {A.}~\bibnamefont {Scherer}}, \emph {et~al.},\ }\bibfield  {title} {\bibinfo {title} {Impacts of decoder latency on utility-scale quantum computer architectures},\ }\href@noop {} {\bibfield  {journal} {\bibinfo  {journal} {arXiv preprint arXiv:2511.10633}\ } (\bibinfo {year} {2025})}\BibitemShut {NoStop}%
\bibitem [{\citenamefont {Tanner}(1981)}]{tanner1981recursive}%
  \BibitemOpen
  \bibfield  {author} {\bibinfo {author} {\bibfnamefont {R.}~\bibnamefont {Tanner}},\ }\bibfield  {title} {\bibinfo {title} {A recursive approach to low complexity codes},\ }\href {https://doi.org/10.1109/TIT.1981.1056404} {\bibfield  {journal} {\bibinfo  {journal} {IEEE Transactions on Information Theory}\ }\textbf {\bibinfo {volume} {27}},\ \bibinfo {pages} {533} (\bibinfo {year} {1981})}\BibitemShut {NoStop}%
\bibitem [{\citenamefont {Hu}\ \emph {et~al.}(2001)\citenamefont {Hu}, \citenamefont {Eleftheriou},\ and\ \citenamefont {Arnold}}]{hu2001progressive}%
  \BibitemOpen
  \bibfield  {author} {\bibinfo {author} {\bibfnamefont {X.-Y.}\ \bibnamefont {Hu}}, \bibinfo {author} {\bibfnamefont {E.}~\bibnamefont {Eleftheriou}},\ and\ \bibinfo {author} {\bibfnamefont {D.-M.}\ \bibnamefont {Arnold}},\ }\bibfield  {title} {\bibinfo {title} {Progressive edge-growth tanner graphs},\ }in\ \href@noop {} {\emph {\bibinfo {booktitle} {GLOBECOM'01. IEEE Global Telecommunications Conference (Cat. No. 01CH37270)}}},\ Vol.~\bibinfo {volume} {2}\ (\bibinfo {organization} {IEEE},\ \bibinfo {year} {2001})\ pp.\ \bibinfo {pages} {995--1001}\BibitemShut {NoStop}%
\bibitem [{\citenamefont {Gottesman}(1997)}]{gottesman1997stabilizer}%
  \BibitemOpen
  \bibfield  {author} {\bibinfo {author} {\bibfnamefont {D.}~\bibnamefont {Gottesman}},\ }\emph {\bibinfo {title} {Stabilizer codes and quantum error correction}},\ \href@noop {} {Ph.D. thesis},\ \bibinfo  {school} {California Institute of Technology} (\bibinfo {year} {1997})\BibitemShut {NoStop}%
\bibitem [{\citenamefont {Poulin}\ and\ \citenamefont {Chung}(2008)}]{poulin2008iterative}%
  \BibitemOpen
  \bibfield  {author} {\bibinfo {author} {\bibfnamefont {D.}~\bibnamefont {Poulin}}\ and\ \bibinfo {author} {\bibfnamefont {Y.}~\bibnamefont {Chung}},\ }\bibfield  {title} {\bibinfo {title} {On the iterative decoding of sparse quantum codes},\ }\href@noop {} {\bibfield  {journal} {\bibinfo  {journal} {Quantum Information \& Computation}\ }\textbf {\bibinfo {volume} {8}},\ \bibinfo {pages} {987} (\bibinfo {year} {2008})}\BibitemShut {NoStop}%
\bibitem [{\citenamefont {Panteleev}\ and\ \citenamefont {Kalachev}(2021)}]{panteleev2021degenerate}%
  \BibitemOpen
  \bibfield  {author} {\bibinfo {author} {\bibfnamefont {P.}~\bibnamefont {Panteleev}}\ and\ \bibinfo {author} {\bibfnamefont {G.}~\bibnamefont {Kalachev}},\ }\bibfield  {title} {\bibinfo {title} {Degenerate quantum {LDPC} codes with good finite length performance},\ }\href@noop {} {\bibfield  {journal} {\bibinfo  {journal} {Quantum}\ }\textbf {\bibinfo {volume} {5}},\ \bibinfo {pages} {585} (\bibinfo {year} {2021})}\BibitemShut {NoStop}%
\bibitem [{\citenamefont {Roffe}\ \emph {et~al.}(2020)\citenamefont {Roffe}, \citenamefont {White}, \citenamefont {Burton},\ and\ \citenamefont {Campbell}}]{roffe_decoding_2020}%
  \BibitemOpen
  \bibfield  {author} {\bibinfo {author} {\bibfnamefont {J.}~\bibnamefont {Roffe}}, \bibinfo {author} {\bibfnamefont {D.~R.}\ \bibnamefont {White}}, \bibinfo {author} {\bibfnamefont {S.}~\bibnamefont {Burton}},\ and\ \bibinfo {author} {\bibfnamefont {E.}~\bibnamefont {Campbell}},\ }\bibfield  {title} {\bibinfo {title} {Decoding across the quantum low-density parity-check code landscape},\ }\href@noop {} {\bibfield  {journal} {\bibinfo  {journal} {Physical Review Research}\ }\textbf {\bibinfo {volume} {2}} (\bibinfo {year} {2020})}\BibitemShut {NoStop}%
\bibitem [{\citenamefont {Roffe}(2022)}]{Roffe_LDPC_Python_tools_2022}%
  \BibitemOpen
  \bibfield  {author} {\bibinfo {author} {\bibfnamefont {J.}~\bibnamefont {Roffe}},\ }\href {https://pypi.org/project/ldpc/} {\bibinfo {title} {{LDPC: Python tools for low density parity check codes}}} (\bibinfo {year} {2022})\BibitemShut {NoStop}%
\bibitem [{\citenamefont {Delfosse}\ \emph {et~al.}(2022)\citenamefont {Delfosse}, \citenamefont {Londe},\ and\ \citenamefont {Beverland}}]{delfosse2022toward}%
  \BibitemOpen
  \bibfield  {author} {\bibinfo {author} {\bibfnamefont {N.}~\bibnamefont {Delfosse}}, \bibinfo {author} {\bibfnamefont {V.}~\bibnamefont {Londe}},\ and\ \bibinfo {author} {\bibfnamefont {M.~E.}\ \bibnamefont {Beverland}},\ }\bibfield  {title} {\bibinfo {title} {Toward a union-find decoder for quantum ldpc codes},\ }\href@noop {} {\bibfield  {journal} {\bibinfo  {journal} {IEEE Transactions on Information Theory}\ }\textbf {\bibinfo {volume} {68}},\ \bibinfo {pages} {3187} (\bibinfo {year} {2022})}\BibitemShut {NoStop}%
\bibitem [{\citenamefont {Wolanski}\ and\ \citenamefont {Barber}(2024)}]{wolanski2024introducing}%
  \BibitemOpen
  \bibfield  {author} {\bibinfo {author} {\bibfnamefont {S.}~\bibnamefont {Wolanski}}\ and\ \bibinfo {author} {\bibfnamefont {B.}~\bibnamefont {Barber}},\ }\bibfield  {title} {\bibinfo {title} {Introducing ambiguity clustering: an accurate and efficient decoder for qldpc codes},\ }in\ \href@noop {} {\emph {\bibinfo {booktitle} {2024 IEEE International Conference on Quantum Computing and Engineering (QCE)}}},\ Vol.~\bibinfo {volume} {2}\ (\bibinfo {organization} {IEEE},\ \bibinfo {year} {2024})\ pp.\ \bibinfo {pages} {402--403}\BibitemShut {NoStop}%
\bibitem [{\citenamefont {Hillmann}\ \emph {et~al.}(2025)\citenamefont {Hillmann}, \citenamefont {Berent}, \citenamefont {Quintavalle}, \citenamefont {Eisert}, \citenamefont {Wille},\ and\ \citenamefont {Roffe}}]{hillmann2024localized}%
  \BibitemOpen
  \bibfield  {author} {\bibinfo {author} {\bibfnamefont {T.}~\bibnamefont {Hillmann}}, \bibinfo {author} {\bibfnamefont {L.}~\bibnamefont {Berent}}, \bibinfo {author} {\bibfnamefont {A.~O.}\ \bibnamefont {Quintavalle}}, \bibinfo {author} {\bibfnamefont {J.}~\bibnamefont {Eisert}}, \bibinfo {author} {\bibfnamefont {R.}~\bibnamefont {Wille}},\ and\ \bibinfo {author} {\bibfnamefont {J.}~\bibnamefont {Roffe}},\ }\bibfield  {title} {\bibinfo {title} {Localized statistics decoding for quantum low-density parity-check codes},\ }\href@noop {} {\bibfield  {journal} {\bibinfo  {journal} {Nature Communications}\ }\textbf {\bibinfo {volume} {16}},\ \bibinfo {pages} {8214} (\bibinfo {year} {2025})}\BibitemShut {NoStop}%
\bibitem [{\citenamefont {Du~Crest}\ \emph {et~al.}(2022)\citenamefont {Du~Crest}, \citenamefont {Mhalla},\ and\ \citenamefont {Savin}}]{du2022stabilizer}%
  \BibitemOpen
  \bibfield  {author} {\bibinfo {author} {\bibfnamefont {J.}~\bibnamefont {Du~Crest}}, \bibinfo {author} {\bibfnamefont {M.}~\bibnamefont {Mhalla}},\ and\ \bibinfo {author} {\bibfnamefont {V.}~\bibnamefont {Savin}},\ }\bibfield  {title} {\bibinfo {title} {Stabilizer inactivation for message-passing decoding of quantum ldpc codes},\ }in\ \href@noop {} {\emph {\bibinfo {booktitle} {2022 IEEE Information Theory Workshop (ITW)}}}\ (\bibinfo {organization} {IEEE},\ \bibinfo {year} {2022})\ pp.\ \bibinfo {pages} {488--493}\BibitemShut {NoStop}%
\bibitem [{\citenamefont {Yao}\ \emph {et~al.}(2024)\citenamefont {Yao}, \citenamefont {Laban}, \citenamefont {H{\"a}ger}, \citenamefont {i~Amat},\ and\ \citenamefont {Pfister}}]{yao2024belief}%
  \BibitemOpen
  \bibfield  {author} {\bibinfo {author} {\bibfnamefont {H.}~\bibnamefont {Yao}}, \bibinfo {author} {\bibfnamefont {W.~A.}\ \bibnamefont {Laban}}, \bibinfo {author} {\bibfnamefont {C.}~\bibnamefont {H{\"a}ger}}, \bibinfo {author} {\bibfnamefont {A.~G.}\ \bibnamefont {i~Amat}},\ and\ \bibinfo {author} {\bibfnamefont {H.~D.}\ \bibnamefont {Pfister}},\ }\bibfield  {title} {\bibinfo {title} {Belief propagation decoding of quantum ldpc codes with guided decimation},\ }in\ \href@noop {} {\emph {\bibinfo {booktitle} {2024 IEEE International Symposium on Information Theory (ISIT)}}}\ (\bibinfo {organization} {IEEE},\ \bibinfo {year} {2024})\ pp.\ \bibinfo {pages} {2478--2483}\BibitemShut {NoStop}%
\bibitem [{\citenamefont {Gong}\ \emph {et~al.}(2024)\citenamefont {Gong}, \citenamefont {Cammerer},\ and\ \citenamefont {Renes}}]{gong2024toward}%
  \BibitemOpen
  \bibfield  {author} {\bibinfo {author} {\bibfnamefont {A.}~\bibnamefont {Gong}}, \bibinfo {author} {\bibfnamefont {S.}~\bibnamefont {Cammerer}},\ and\ \bibinfo {author} {\bibfnamefont {J.~M.}\ \bibnamefont {Renes}},\ }\bibfield  {title} {\bibinfo {title} {Toward low-latency iterative decoding of {QLDPC} codes under circuit-level noise},\ }\href@noop {} {\bibfield  {journal} {\bibinfo  {journal} {arXiv:2403.18901}\ } (\bibinfo {year} {2024})}\BibitemShut {NoStop}%
\bibitem [{\citenamefont {iOlius}\ \emph {et~al.}(2024)\citenamefont {iOlius}, \citenamefont {Martinez}, \citenamefont {Roffe},\ and\ \citenamefont {Martinez}}]{iolius2024almost}%
  \BibitemOpen
  \bibfield  {author} {\bibinfo {author} {\bibfnamefont {A.~d.}\ \bibnamefont {iOlius}}, \bibinfo {author} {\bibfnamefont {I.~E.}\ \bibnamefont {Martinez}}, \bibinfo {author} {\bibfnamefont {J.}~\bibnamefont {Roffe}},\ and\ \bibinfo {author} {\bibfnamefont {J.~E.}\ \bibnamefont {Martinez}},\ }\bibfield  {title} {\bibinfo {title} {An almost-linear time decoding algorithm for quantum ldpc codes under circuit-level noise},\ }\href@noop {} {\bibfield  {journal} {\bibinfo  {journal} {arXiv preprint arXiv:2409.01440}\ } (\bibinfo {year} {2024})}\BibitemShut {NoStop}%
\bibitem [{\citenamefont {Koutsioumpas}\ \emph {et~al.}(2025)\citenamefont {Koutsioumpas}, \citenamefont {Sayginel}, \citenamefont {Webster},\ and\ \citenamefont {Browne}}]{koutsioumpas2025automorphism}%
  \BibitemOpen
  \bibfield  {author} {\bibinfo {author} {\bibfnamefont {S.}~\bibnamefont {Koutsioumpas}}, \bibinfo {author} {\bibfnamefont {H.}~\bibnamefont {Sayginel}}, \bibinfo {author} {\bibfnamefont {M.}~\bibnamefont {Webster}},\ and\ \bibinfo {author} {\bibfnamefont {D.~E.}\ \bibnamefont {Browne}},\ }\bibfield  {title} {\bibinfo {title} {Automorphism ensemble decoding of quantum ldpc codes},\ }\href@noop {} {\bibfield  {journal} {\bibinfo  {journal} {arXiv preprint arXiv:2503.01738}\ } (\bibinfo {year} {2025})}\BibitemShut {NoStop}%
\bibitem [{\citenamefont {Wang}\ \emph {et~al.}(2026{\natexlab{a}})\citenamefont {Wang}, \citenamefont {Li},\ and\ \citenamefont {Mueller}}]{wang2026fully}%
  \BibitemOpen
  \bibfield  {author} {\bibinfo {author} {\bibfnamefont {M.}~\bibnamefont {Wang}}, \bibinfo {author} {\bibfnamefont {A.}~\bibnamefont {Li}},\ and\ \bibinfo {author} {\bibfnamefont {F.}~\bibnamefont {Mueller}},\ }\bibfield  {title} {\bibinfo {title} {Fully parallelized bp decoding for quantum ldpc codes can outperform bp-osd},\ }in\ \href@noop {} {\emph {\bibinfo {booktitle} {2026 IEEE International Symposium on High Performance Computer Architecture (HPCA)}}}\ (\bibinfo {organization} {IEEE},\ \bibinfo {year} {2026})\ pp.\ \bibinfo {pages} {1--14}\BibitemShut {NoStop}%
\bibitem [{\citenamefont {Pradhan}\ \emph {et~al.}(2026)\citenamefont {Pradhan}, \citenamefont {Raveendran}, \citenamefont {Declercq},\ and\ \citenamefont {Vasi{\'c}}}]{pradhan2026fast}%
  \BibitemOpen
  \bibfield  {author} {\bibinfo {author} {\bibfnamefont {A.~K.}\ \bibnamefont {Pradhan}}, \bibinfo {author} {\bibfnamefont {N.}~\bibnamefont {Raveendran}}, \bibinfo {author} {\bibfnamefont {D.}~\bibnamefont {Declercq}},\ and\ \bibinfo {author} {\bibfnamefont {B.}~\bibnamefont {Vasi{\'c}}},\ }\href@noop {} {\bibinfo {title} {Fast fault-tolerant decoders for hypergraph product and lifted-product codes}} (\bibinfo {year} {2026}),\ \bibinfo {note} {arXiv:2608.31040}\BibitemShut {NoStop}%
\bibitem [{\citenamefont {Ott}\ \emph {et~al.}(2025)\citenamefont {Ott}, \citenamefont {Het{\'e}nyi},\ and\ \citenamefont {Beverland}}]{ott2025decision}%
  \BibitemOpen
  \bibfield  {author} {\bibinfo {author} {\bibfnamefont {K.~R.}\ \bibnamefont {Ott}}, \bibinfo {author} {\bibfnamefont {B.}~\bibnamefont {Het{\'e}nyi}},\ and\ \bibinfo {author} {\bibfnamefont {M.~E.}\ \bibnamefont {Beverland}},\ }\href@noop {} {\bibinfo {title} {Decision-tree decoders for general quantum {LDPC} codes}} (\bibinfo {year} {2025}),\ \bibinfo {note} {arXiv:2502.16408}\BibitemShut {NoStop}%
\bibitem [{\citenamefont {Wu}\ \emph {et~al.}(2025)\citenamefont {Wu}, \citenamefont {Li}, \citenamefont {Chang}, \citenamefont {Puri},\ and\ \citenamefont {Zhong}}]{wu2025minimum}%
  \BibitemOpen
  \bibfield  {author} {\bibinfo {author} {\bibfnamefont {Y.}~\bibnamefont {Wu}}, \bibinfo {author} {\bibfnamefont {B.}~\bibnamefont {Li}}, \bibinfo {author} {\bibfnamefont {K.}~\bibnamefont {Chang}}, \bibinfo {author} {\bibfnamefont {S.}~\bibnamefont {Puri}},\ and\ \bibinfo {author} {\bibfnamefont {L.}~\bibnamefont {Zhong}},\ }\bibfield  {title} {\bibinfo {title} {Minimum-weight parity factor decoder for quantum error correction},\ }\href@noop {} {\bibfield  {journal} {\bibinfo  {journal} {arXiv preprint arXiv:2508.04969}\ } (\bibinfo {year} {2025})}\BibitemShut {NoStop}%
\bibitem [{\citenamefont {Beni}\ \emph {et~al.}(2025)\citenamefont {Beni}, \citenamefont {Higgott},\ and\ \citenamefont {Shutty}}]{beni2025tesseract}%
  \BibitemOpen
  \bibfield  {author} {\bibinfo {author} {\bibfnamefont {L.~A.}\ \bibnamefont {Beni}}, \bibinfo {author} {\bibfnamefont {O.}~\bibnamefont {Higgott}},\ and\ \bibinfo {author} {\bibfnamefont {N.}~\bibnamefont {Shutty}},\ }\bibfield  {title} {\bibinfo {title} {Tesseract: A search-based decoder for quantum error correction},\ }\href@noop {} {\bibfield  {journal} {\bibinfo  {journal} {arXiv preprint arXiv:2503.10988}\ } (\bibinfo {year} {2025})}\BibitemShut {NoStop}%
\bibitem [{\citenamefont {Valentini}\ \emph {et~al.}(2026)\citenamefont {Valentini}, \citenamefont {Forlivesi}, \citenamefont {Talarico},\ and\ \citenamefont {Chiani}}]{valentini2026restart}%
  \BibitemOpen
  \bibfield  {author} {\bibinfo {author} {\bibfnamefont {L.}~\bibnamefont {Valentini}}, \bibinfo {author} {\bibfnamefont {D.}~\bibnamefont {Forlivesi}}, \bibinfo {author} {\bibfnamefont {A.}~\bibnamefont {Talarico}},\ and\ \bibinfo {author} {\bibfnamefont {M.}~\bibnamefont {Chiani}},\ }\bibfield  {title} {\bibinfo {title} {Restart belief: A general quantum ldpc decoder},\ }\href@noop {} {\bibfield  {journal} {\bibinfo  {journal} {IEEE Communications Letters}\ } (\bibinfo {year} {2026})}\BibitemShut {NoStop}%
\bibitem [{\citenamefont {Leverrier}\ and\ \citenamefont {Urbanke}(2026)}]{leverrier2026approximating}%
  \BibitemOpen
  \bibfield  {author} {\bibinfo {author} {\bibfnamefont {A.}~\bibnamefont {Leverrier}}\ and\ \bibinfo {author} {\bibfnamefont {R.}~\bibnamefont {Urbanke}},\ }\href@noop {} {\bibinfo {title} {Approximating optimal decoding of quantum {LDPC} codes with narrow frontiers}} (\bibinfo {year} {2026}),\ \bibinfo {note} {arXiv:2606.20513}\BibitemShut {NoStop}%
\bibitem [{\citenamefont {M{\"u}ller}\ \emph {et~al.}(2025)\citenamefont {M{\"u}ller}, \citenamefont {Alexander}, \citenamefont {Beverland}, \citenamefont {B{\"u}hler}, \citenamefont {Johnson}, \citenamefont {Maurer},\ and\ \citenamefont {Vandeth}}]{muller2025improved}%
  \BibitemOpen
  \bibfield  {author} {\bibinfo {author} {\bibfnamefont {T.}~\bibnamefont {M{\"u}ller}}, \bibinfo {author} {\bibfnamefont {T.}~\bibnamefont {Alexander}}, \bibinfo {author} {\bibfnamefont {M.~E.}\ \bibnamefont {Beverland}}, \bibinfo {author} {\bibfnamefont {M.}~\bibnamefont {B{\"u}hler}}, \bibinfo {author} {\bibfnamefont {B.~R.}\ \bibnamefont {Johnson}}, \bibinfo {author} {\bibfnamefont {T.}~\bibnamefont {Maurer}},\ and\ \bibinfo {author} {\bibfnamefont {D.}~\bibnamefont {Vandeth}},\ }\href@noop {} {\bibinfo {title} {Improved belief propagation is sufficient for real-time decoding of quantum memory}} (\bibinfo {year} {2025}),\ \bibinfo {note} {arXiv:2506.01779}\BibitemShut {NoStop}%
\bibitem [{\citenamefont {Maurer}\ \emph {et~al.}(2025)\citenamefont {Maurer}, \citenamefont {B{\"u}hler}, \citenamefont {Kr{\"o}ner}, \citenamefont {Haverkamp}, \citenamefont {M{\"u}ller}, \citenamefont {Vandeth},\ and\ \citenamefont {Johnson}}]{maurer2025real}%
  \BibitemOpen
  \bibfield  {author} {\bibinfo {author} {\bibfnamefont {T.}~\bibnamefont {Maurer}}, \bibinfo {author} {\bibfnamefont {M.}~\bibnamefont {B{\"u}hler}}, \bibinfo {author} {\bibfnamefont {M.}~\bibnamefont {Kr{\"o}ner}}, \bibinfo {author} {\bibfnamefont {F.}~\bibnamefont {Haverkamp}}, \bibinfo {author} {\bibfnamefont {T.}~\bibnamefont {M{\"u}ller}}, \bibinfo {author} {\bibfnamefont {D.}~\bibnamefont {Vandeth}},\ and\ \bibinfo {author} {\bibfnamefont {B.~R.}\ \bibnamefont {Johnson}},\ }\bibfield  {title} {\bibinfo {title} {Real-time decoding of the gross code memory with {FPGAs}},\ }\href@noop {} {\bibfield  {journal} {\bibinfo  {journal} {arXiv preprint arXiv:2510.21600}\ } (\bibinfo {year} {2025})}\BibitemShut {NoStop}%
\bibitem [{\citenamefont {Maurya}\ \emph {et~al.}(2025)\citenamefont {Maurya}, \citenamefont {Maurer}, \citenamefont {B{\"u}hler}, \citenamefont {Vandeth},\ and\ \citenamefont {Beverland}}]{maurya2025fpga}%
  \BibitemOpen
  \bibfield  {author} {\bibinfo {author} {\bibfnamefont {S.}~\bibnamefont {Maurya}}, \bibinfo {author} {\bibfnamefont {T.}~\bibnamefont {Maurer}}, \bibinfo {author} {\bibfnamefont {M.}~\bibnamefont {B{\"u}hler}}, \bibinfo {author} {\bibfnamefont {D.}~\bibnamefont {Vandeth}},\ and\ \bibinfo {author} {\bibfnamefont {M.~E.}\ \bibnamefont {Beverland}},\ }\bibfield  {title} {\bibinfo {title} {{FPGA-tailored} algorithms for real-time decoding of quantum ldpc codes},\ }\href@noop {} {\bibfield  {journal} {\bibinfo  {journal} {arXiv preprint arXiv:2511.21660}\ } (\bibinfo {year} {2025})}\BibitemShut {NoStop}%
\bibitem [{\citenamefont {Markidis}\ \emph {et~al.}(2026)\citenamefont {Markidis}, \citenamefont {Pennati},\ and\ \citenamefont {Peng}}]{markidis2026engineering}%
  \BibitemOpen
  \bibfield  {author} {\bibinfo {author} {\bibfnamefont {S.}~\bibnamefont {Markidis}}, \bibinfo {author} {\bibfnamefont {L.}~\bibnamefont {Pennati}},\ and\ \bibinfo {author} {\bibfnamefont {I.}~\bibnamefont {Peng}},\ }\href@noop {} {\bibinfo {title} {Engineering low-latency, high-throughput quantum decoders on nvidia gpus}} (\bibinfo {year} {2026}),\ \bibinfo {note} {arXiv:2610.28867}\BibitemShut {NoStop}%
\bibitem [{\citenamefont {Yoder}\ \emph {et~al.}(2025)\citenamefont {Yoder}, \citenamefont {Schoute}, \citenamefont {Rall}, \citenamefont {Pritchett}, \citenamefont {Gambetta}, \citenamefont {Cross}, \citenamefont {Carroll},\ and\ \citenamefont {Beverland}}]{yoder2025tour}%
  \BibitemOpen
  \bibfield  {author} {\bibinfo {author} {\bibfnamefont {T.~J.}\ \bibnamefont {Yoder}}, \bibinfo {author} {\bibfnamefont {E.}~\bibnamefont {Schoute}}, \bibinfo {author} {\bibfnamefont {P.}~\bibnamefont {Rall}}, \bibinfo {author} {\bibfnamefont {E.}~\bibnamefont {Pritchett}}, \bibinfo {author} {\bibfnamefont {J.~M.}\ \bibnamefont {Gambetta}}, \bibinfo {author} {\bibfnamefont {A.~W.}\ \bibnamefont {Cross}}, \bibinfo {author} {\bibfnamefont {M.}~\bibnamefont {Carroll}},\ and\ \bibinfo {author} {\bibfnamefont {M.~E.}\ \bibnamefont {Beverland}},\ }\href@noop {} {\bibinfo {title} {Tour de gross: A modular quantum computer based on bivariate bicycle codes}} (\bibinfo {year} {2025}),\ \bibinfo {note} {arXiv:2506.03094}\BibitemShut {NoStop}%
\bibitem [{\citenamefont {Hong}(2026)}]{hong2026quantum}%
  \BibitemOpen
  \bibfield  {author} {\bibinfo {author} {\bibfnamefont {Y.}~\bibnamefont {Hong}},\ }\href@noop {} {\bibinfo {title} {Quantum {LDPC} codes with design rate 1/5 and good performance below 1000 physical qubits}} (\bibinfo {year} {2026}),\ \bibinfo {note} {arXiv:2607.27644}\BibitemShut {NoStop}%
\bibitem [{\citenamefont {Maan}\ \emph {et~al.}(2026)\citenamefont {Maan}, \citenamefont {Garcia~Herrero}, \citenamefont {Paler},\ and\ \citenamefont {Savin}}]{maan2026decoding}%
  \BibitemOpen
  \bibfield  {author} {\bibinfo {author} {\bibfnamefont {A.~S.}\ \bibnamefont {Maan}}, \bibinfo {author} {\bibfnamefont {F.~M.}\ \bibnamefont {Garcia~Herrero}}, \bibinfo {author} {\bibfnamefont {A.}~\bibnamefont {Paler}},\ and\ \bibinfo {author} {\bibfnamefont {V.}~\bibnamefont {Savin}},\ }\bibfield  {title} {\bibinfo {title} {Decoding correlated errors in quantum {LDPC} codes},\ }\href@noop {} {\bibfield  {journal} {\bibinfo  {journal} {Nature Communications}\ }\textbf {\bibinfo {volume} {17}},\ \bibinfo {pages} {3965} (\bibinfo {year} {2026})}\BibitemShut {NoStop}%
\bibitem [{\citenamefont {Ye}\ \emph {et~al.}(2026{\natexlab{a}})\citenamefont {Ye}, \citenamefont {Wecker},\ and\ \citenamefont {Delfosse}}]{ye2026beam}%
  \BibitemOpen
  \bibfield  {author} {\bibinfo {author} {\bibfnamefont {M.}~\bibnamefont {Ye}}, \bibinfo {author} {\bibfnamefont {D.}~\bibnamefont {Wecker}},\ and\ \bibinfo {author} {\bibfnamefont {N.}~\bibnamefont {Delfosse}},\ }\bibfield  {title} {\bibinfo {title} {Beam search decoder for quantum low-density parity-check codes},\ }\href@noop {} {\bibfield  {journal} {\bibinfo  {journal} {PRX Quantum}\ }\textbf {\bibinfo {volume} {7}},\ \bibinfo {pages} {033002} (\bibinfo {year} {2026}{\natexlab{a}})}\BibitemShut {NoStop}%
\bibitem [{\citenamefont {Tham}\ \emph {et~al.}(2026)\citenamefont {Tham}, \citenamefont {Goldman}, \citenamefont {Debnath}, \citenamefont {Patel}, \citenamefont {Rasheed}, \citenamefont {Nielsen}, \citenamefont {Boguslawski}, \citenamefont {Mizrahi}, \citenamefont {Nguyen}, \citenamefont {Saraladevi}, \citenamefont {Tripier}, \citenamefont {West}, \citenamefont {Ye}, \citenamefont {Pisenti}, \citenamefont {Wright}, \citenamefont {Gamble},\ and\ \citenamefont {Delfosse}}]{tham2026breakeven}%
  \BibitemOpen
  \bibfield  {author} {\bibinfo {author} {\bibfnamefont {E.}~\bibnamefont {Tham}}, \bibinfo {author} {\bibfnamefont {M.~L.}\ \bibnamefont {Goldman}}, \bibinfo {author} {\bibfnamefont {S.}~\bibnamefont {Debnath}}, \bibinfo {author} {\bibfnamefont {A.~N.}\ \bibnamefont {Patel}}, \bibinfo {author} {\bibfnamefont {M.}~\bibnamefont {Rasheed}}, \bibinfo {author} {\bibfnamefont {E.}~\bibnamefont {Nielsen}}, \bibinfo {author} {\bibfnamefont {M.}~\bibnamefont {Boguslawski}}, \bibinfo {author} {\bibfnamefont {J.}~\bibnamefont {Mizrahi}}, \bibinfo {author} {\bibfnamefont {J.}~\bibnamefont {Nguyen}}, \bibinfo {author} {\bibfnamefont {J.}~\bibnamefont {Saraladevi}}, \bibinfo {author} {\bibfnamefont {F.}~\bibnamefont {Tripier}}, \bibinfo {author} {\bibfnamefont {A.}~\bibnamefont {West}}, \bibinfo {author} {\bibfnamefont {M.}~\bibnamefont {Ye}}, \bibinfo {author} {\bibfnamefont {N.}~\bibnamefont {Pisenti}}, \bibinfo {author} {\bibfnamefont {K.}~\bibnamefont {Wright}}, \bibinfo {author} {\bibfnamefont {J.}~\bibnamefont
  {Gamble}},\ and\ \bibinfo {author} {\bibfnamefont {N.}~\bibnamefont {Delfosse}},\ }\href@noop {} {\bibinfo {title} {Breakeven demonstration of quantum low-density parity-check codes}} (\bibinfo {year} {2026}),\ \bibinfo {note} {arXiv:2606.06455}\BibitemShut {NoStop}%
\bibitem [{\citenamefont {Hines}\ \emph {et~al.}(2026)\citenamefont {Hines}, \citenamefont {Ostrove}, \citenamefont {Rudinger}, \citenamefont {Seritan}, \citenamefont {Young}, \citenamefont {Blume-Kohout},\ and\ \citenamefont {Proctor}}]{hines2026simulating}%
  \BibitemOpen
  \bibfield  {author} {\bibinfo {author} {\bibfnamefont {J.}~\bibnamefont {Hines}}, \bibinfo {author} {\bibfnamefont {C.}~\bibnamefont {Ostrove}}, \bibinfo {author} {\bibfnamefont {K.}~\bibnamefont {Rudinger}}, \bibinfo {author} {\bibfnamefont {S.}~\bibnamefont {Seritan}}, \bibinfo {author} {\bibfnamefont {K.}~\bibnamefont {Young}}, \bibinfo {author} {\bibfnamefont {R.}~\bibnamefont {Blume-Kohout}},\ and\ \bibinfo {author} {\bibfnamefont {T.}~\bibnamefont {Proctor}},\ }\href@noop {} {\bibinfo {title} {Simulating quantum error correction beyond pauli stochastic errors}} (\bibinfo {year} {2026}),\ \bibinfo {note} {arXiv:2603.18457}\BibitemShut {NoStop}%
\bibitem [{\citenamefont {Tripier}\ \emph {et~al.}(2026)\citenamefont {Tripier}, \citenamefont {Chung}, \citenamefont {Young}, \citenamefont {Alam}, \citenamefont {Bjork}, \citenamefont {Brodutch}, \citenamefont {Buessen}, \citenamefont {Coble}, \citenamefont {Dellaert}, \citenamefont {Maslov} \emph {et~al.}}]{tripier2026fault}%
  \BibitemOpen
  \bibfield  {author} {\bibinfo {author} {\bibfnamefont {F.}~\bibnamefont {Tripier}}, \bibinfo {author} {\bibfnamefont {W.~C.}\ \bibnamefont {Chung}}, \bibinfo {author} {\bibfnamefont {J.}~\bibnamefont {Young}}, \bibinfo {author} {\bibfnamefont {S.}~\bibnamefont {Alam}}, \bibinfo {author} {\bibfnamefont {B.}~\bibnamefont {Bjork}}, \bibinfo {author} {\bibfnamefont {A.}~\bibnamefont {Brodutch}}, \bibinfo {author} {\bibfnamefont {F.~L.}\ \bibnamefont {Buessen}}, \bibinfo {author} {\bibfnamefont {N.~J.}\ \bibnamefont {Coble}}, \bibinfo {author} {\bibfnamefont {T.}~\bibnamefont {Dellaert}}, \bibinfo {author} {\bibfnamefont {D.}~\bibnamefont {Maslov}}, \emph {et~al.},\ }\href@noop {} {\bibinfo {title} {Fault-tolerant quantum computing with trapped ions: The walking cat architecture}} (\bibinfo {year} {2026}),\ \bibinfo {note} {arXiv:2604.19481}\BibitemShut {NoStop}%
\bibitem [{\citenamefont {Ye}\ \emph {et~al.}(2026{\natexlab{b}})\citenamefont {Ye}, \citenamefont {Maksymov},\ and\ \citenamefont {Delfosse}}]{ye2026real}%
  \BibitemOpen
  \bibfield  {author} {\bibinfo {author} {\bibfnamefont {M.}~\bibnamefont {Ye}}, \bibinfo {author} {\bibfnamefont {A.}~\bibnamefont {Maksymov}},\ and\ \bibinfo {author} {\bibfnamefont {N.}~\bibnamefont {Delfosse}},\ }\href@noop {} {\bibinfo {title} {Real-time decoder for a {MegaQuOp} quantum computer using a single {CPU}}} (\bibinfo {year} {2026}{\natexlab{b}}),\ \bibinfo {note} {arXiv:2608.25027}\BibitemShut {NoStop}%
\bibitem [{\citenamefont {Taghipour}\ \emph {et~al.}(2026)\citenamefont {Taghipour}, \citenamefont {Chytas},\ and\ \citenamefont {Vasi{\'c}}}]{taghipour2026multistage}%
  \BibitemOpen
  \bibfield  {author} {\bibinfo {author} {\bibfnamefont {M.}~\bibnamefont {Taghipour}}, \bibinfo {author} {\bibfnamefont {D.}~\bibnamefont {Chytas}},\ and\ \bibinfo {author} {\bibfnamefont {B.}~\bibnamefont {Vasi{\'c}}},\ }\href@noop {} {\bibinfo {title} {Multistage rewinding decoder for {QLDPC} codes}} (\bibinfo {year} {2026}),\ \bibinfo {note} {arXiv:2608.07783}\BibitemShut {NoStop}%
\bibitem [{\citenamefont {Aydin}\ \emph {et~al.}(2026)\citenamefont {Aydin}, \citenamefont {Tham}, \citenamefont {Delfosse},\ and\ \citenamefont {Ye}}]{aydin2026soft}%
  \BibitemOpen
  \bibfield  {author} {\bibinfo {author} {\bibfnamefont {A.}~\bibnamefont {Aydin}}, \bibinfo {author} {\bibfnamefont {E.}~\bibnamefont {Tham}}, \bibinfo {author} {\bibfnamefont {N.}~\bibnamefont {Delfosse}},\ and\ \bibinfo {author} {\bibfnamefont {M.}~\bibnamefont {Ye}},\ }\href@noop {} {\bibinfo {title} {Soft decoding for quantum {LDPC} codes with experimental validation}} (\bibinfo {year} {2026}),\ \bibinfo {note} {arXiv:2609.26958}\BibitemShut {NoStop}%
\bibitem [{pri(2026)}]{priortilt_github}%
  \BibitemOpen
  \href@noop {} {\bibinfo {title} {Source code of {PriorTilt} decoder}},\ \bibinfo {howpublished} {\url{https://github.com/ionq-publications/PriorTiltDecoder}} (\bibinfo {year} {2026})\BibitemShut {NoStop}%
\bibitem [{\citenamefont {Wang}\ \emph {et~al.}(2026{\natexlab{b}})\citenamefont {Wang}, \citenamefont {Zhao}, \citenamefont {Li},\ and\ \citenamefont {Mueller}}]{wang2026gpu}%
  \BibitemOpen
  \bibfield  {author} {\bibinfo {author} {\bibfnamefont {M.}~\bibnamefont {Wang}}, \bibinfo {author} {\bibfnamefont {Y.}~\bibnamefont {Zhao}}, \bibinfo {author} {\bibfnamefont {A.}~\bibnamefont {Li}},\ and\ \bibinfo {author} {\bibfnamefont {F.}~\bibnamefont {Mueller}},\ }\href@noop {} {\bibinfo {title} {{GPU}-based parallel {BP} decoding for quantum {LDPC} memory}} (\bibinfo {year} {2026}{\natexlab{b}}),\ \bibinfo {note} {arXiv:2610.13879}\BibitemShut {NoStop}%
\bibitem [{\citenamefont {Gidney}(2021)}]{gidney2021stim}%
  \BibitemOpen
  \bibfield  {author} {\bibinfo {author} {\bibfnamefont {C.}~\bibnamefont {Gidney}},\ }\bibfield  {title} {\bibinfo {title} {Stim: a fast stabilizer circuit simulator},\ }\href {https://doi.org/10.22331/q-2021-07-06-497} {\bibfield  {journal} {\bibinfo  {journal} {Quantum}\ }\textbf {\bibinfo {volume} {5}},\ \bibinfo {pages} {497} (\bibinfo {year} {2021})}\BibitemShut {NoStop}%
\end{thebibliography}%

\end{document}